\documentclass[secnumarabic,amssymb, nobibnotes, aps, prd,onecolumn]{revtex4-2}
\usepackage{graphicx,epsfig,youngtab}

\usepackage{xcolor}
\usepackage{bbold}
\usepackage{amsmath}
\usepackage{hyperref}
\usepackage{amsfonts}
\expandafter\let\csname equation*\endcsname\relax 
\expandafter\let\csname endequation*\endcsname\relax 
\usepackage{bm}
\usepackage{lmodern}
\usepackage{moresize}
\usepackage{natbib}
\usepackage{tikz}
\usetikzlibrary{shapes.symbols,decorations.pathmorphing}
\usepackage{cabin}
\usepackage{pgfplots}
\pgfplotsset{compat = newest}

\def\q {\bm{q}}
\def\x {\bm{x}}
\def\r {\bm{r}}
\def\d {{\rm{d}}}

\def\k {\bm{k}}
\def\x {\bm{x}}

\newcommand{\HH}{\mathcal{H}\,}

\newcommand{\<}{\langle}
\renewcommand{\>}{\rangle}

\def\jnlref#1{{\rm#1}}
\def\aj{\ref@jnl{AJ}}                   
\def\actaa{\ref@jnl{Acta Astron.}}      
\def\araa{\ref@jnl{ARA\&A}}             
\def\apj{\ref@jnl{ApJ}}                 
\def\apjl{\ref@jnl{ApJ}}                
\def\apjs{\ref@jnl{ApJS}}               
\def\ao{\ref@jnl{Appl.~Opt.}}           
\def\apss{\ref@jnl{Ap\&SS}}             
\def\aap{\ref@jnl{A\&A}}                
\def\aapr{\ref@jnl{A\&A~Rev.}}          
\def\aaps{\ref@jnl{A\&AS}}              
\def\azh{\ref@jnl{AZh}}                 
\def\baas{\ref@jnl{BAAS}}               
\def\mnras{\jnlref{MNRAS}}

\begin{document}

         \title{\boldmath An essential building block for cosmological zoom-in perturbation theory}

\author{Obinna Umeh }
\affiliation{${}^{1}$Institute of Cosmology \& Gravitation  University of Portsmouth, Portsmouth PO1 3FX, United Kingdom}
\affiliation{${}^{2}$Department of Physics, University of the Western Cape, Cape Town 7535, South Africa}
\affiliation{${}^{3}$Hierarchical Intelligence Lab, Chalisco Phase II, Warri, Delta state, Nigeria}
\email{obinna.umeh@port.ac.uk}
\date{\today}

\begin{abstract}

The evolution of large-scale structure within the standard model of cosmology is well-posed only up to the onset of shell crossing, where particle trajectories appear to intersect. Beyond this point, the evolution equations become non-predictive and perturbative approaches break down.
We show that in General Relativity,  a matter horizon forms before caustics develop for a well-defined initial over-density on an expanding FLRW spacetime.  The matter horizon was first identified by Ellis and Stoeger in 2010  as a dynamical causal boundary that encloses a sub-region of spacetime where structure formation actually takes place. 
We construct a multi-scale hierarchical framework for the propagation of the geodesic congruences that avoids the shell-crossing singularity by cutting the spacetime at the matter horizon and glueing to another spacetime with opposite orientation. 
We identify a relationship between the multi‑scale hierarchical framework and the cosmological zoom-in N-body simulation approach, and relate the local sub-region that decoupled from the Hubble flow to the region of interest in the cosmological zoom-in N-body simulation approach.  Most importantly, the multi‑scale hierarchical framework provides a more robust way of implementing the boundary conditions, which could benefit the cosmological zoom-in N-body simulation approach.

\end{abstract}

\maketitle
\DeclareGraphicsRule{.wmf}{bmp}{jpg}{}{}

\tableofcontents
\maketitle

\section{Introduction}\label{sec:intro}

The clustering of matter on small scales remains a fundamental open problem in cosmology~\cite{Umeh:2021xqm,Ivanov:2022mrd,Paul:2023yrr,Padmanabhan:2023hfr}. Despite significant advances in our understanding of structure formation on large scales~\cite{Bernardeau:2001qr,Umeh:2015gza,Umeh:2016nuh,Jolicoeur:2017eyi,Jolicoeur:2017nyt,Jolicoeur:2018blf,Koyama:2018ttg,Clarkson:2018dwn,Umeh:2019qyd,Umeh:2019jqg,Maartens:2020jzf}, the current theoretical tools, for example, the effective field theory of large-scale structure (EFTofLSS) are limited to quasi-linear scales, typically up to wavenumbers of $ k \sim  0.4h / \rm{Mpc}$ \cite{Umeh:2015gza,Umeh:2016thy,Ivanov:2022mrd}. 
Vlasov perturbation theory has also been developed, it goes beyond the traditional fluid limit to capture the impact of higher order cumulants ~\cite{Garny:2022tlk,Garny:2022kbk,Garny:2025zlq}.
On smaller, non-linear scales, the highly complex nature of gravitational interactions has led to reliance on empirical models, such as the halo occupation distribution (HOD) \cite{Sheth:1999mn,Sheth:2001dp,Cooray:2002dia} and subhalo abundance matching (SHAM)\cite{Masaki:2022zte}, often combined with N-body simulations \cite{Nishimichi:2011jm,Chaves-Montero:2015iga}. These models, while powerful, remain heuristic in nature and rely on approximations that overlook the fundamental aspects of General relativity (GR) on small scales.

A major obstacle in developing a consistent analytical framework for probing matter distributions on small scales is that gravitational collapse drives density contrasts to large values after a certain scale.  This is associated with the formation of a gravitational focusing singularity(shell crossing singularity) at finite time, which eventually leads to a loss of predictability~\cite{Bernardeau:2001qr,Umeh:2015gza,Umeh:2016thy}. 
 In cosmological N-body simulations, this manifests as an explosion in the cost of evaluating gravitational interaction at high resolution within a large cosmological volume~\cite{Dehnen:2000nh,Zhang:2018nqh}. 
The cosmological zoom-in N-body simulation solves the problem of the explosion in computational cost by running two sets of simulations.  Firstly, a large volume  N-body simulation is run at low resolution, then followed by a high resolution of a small region of interest in the presence of the large volume low resolution N-body simulation~\cite{Hahn:2015sia,Ondaro-Mallea:2023qat}.  This has become a go-to approach for simulating clustering of matter on small scales~\cite{AGORA:2021jss,Wetzel:2022man,Nadler:2022dvo}.

The key driver of the explosion in computational cost is the dynamical timescale, which determines the cost of time integration. In the Newtonian limit, the dynamical timescale is inversely proportional to the square root of the matter density. At a caustic, the Newtonian fluid description predicts that the gravitational collapse happens infinitely fast(instantaneously), which is unphysical. 
In general relativity, however, a matter horizon forms first before a caustic. The caustic is only a consequence of extending a one-parameter family of geodesics that has ceased to be proper time maximising beyond the matter horizon. 
Since general relativity allows the freedom to choose an appropriate affine parameter for the geodesic and coordinate system orientation for the spacetime, the singularity avoidance by surgery is a natural, well-motivated physical technique~\cite{Einstein:PhysRev.48.73}..
 The gravitational focusing singularity avoidance leads to a description of the physical reality with two sheets of spacetime which is equivalent to simulating the physical relativity with both large low-resolution box and a small high-resolution box~\cite{Wetzel:2022man}.
We argue in section \ref{sec:GR} and \ref{sec:FLRW_discrete_symmetry} that gravitational focusing singularity avoidance by surgery is equivalent to how the cosmological zoom-in N-body simulation avoids exploding computational cost by running two separate simulations at different resolutions with suitable boundary conditions. 
In our approach, the choice of the region of interest is no longer subjective; rather, it is dynamically defined by the matter horizon introduced by Ellis and Stoeger in 2010~\cite{Ellis:2010fr}.

In General relativity, causal influences propagate along non-spacelike geodesics: null and time-like geodesics~\cite{HawkingandEllis:1973lsss.book}.  
However, the majority of the attention is paid to the horizons formed by null geodesics, that is, the particle horizon and the event horizons~\cite{Ellisbook:2012}.  These are global horizons associated with the global properties of spacetime.  There are also dynamical horizons, such as the apparent horizon, which are associated with the local properties of spacetime~\cite{Penrose:1969pc,Ashtekar:2003hk}.
Matter horizon is a dynamical horizon formed by time-like geodesics \cite{Ellis:2010fr,Umeh:2022kqs,Umeh:2022prn,Umeh:2022hab}. 
On small scales, scalar perturbations (massive particles) dominate gravitational interactions; therefore, a more pertinent concept in probing matter clustering on small scales is the matter horizon. Unfortunately, its impact on clustering on small scales is largely unknown. Little or no attention has been paid to this crucial property of spacetime. This is the gap we plan to fill.
Essentially, we show that taking the matter horizon formed by massive particles into account allows us to avoid the shell crossing singularity, leading to a multi-scale universe where the internal structure of a gravitationally bound system is described as a separate universe at higher resolution.

This paper is structured as follows: in section \ref{sec:GR}, we review the cosmological zoom-in N-body simulation and provide a consistent definition of matter horizon in general relativity. We describe how the description of physical reality naturally requires more than one sheet of spacetime in general relativity in section \ref{sec:FLRW_discrete_symmetry} . 
We conclude in section \ref{sec:conc}.
We use the  Planck CMB constraint on cosmological parameters for quantitative estimates. We use:  $h = 0.674$ for the dimensionless Hubble parameter, $\Omega_{\rm b} = 0.0493$ for baryon density parameter,  $\Omega_{\rm{cdm}} = 0.264$ for the dark matter density parameter,  $\Omega_{\rm m} = \Omega_{\rm{cdm}} + \Omega_{\rm b}$ for the  matter density parameter,  $n_{\rm s} = 0.9608$ for spectral index,  and  $A_{\rm s} = 2.198 \times 10^{-9}$ for the amplitude of the primordial  curvature perturbation~\cite{Aghanim:2018eyx}. 
The small English alphabets from $a-e$ denote the full spacetime indices, while $i$ and $j$ denote the spatial indices.  The capital English alphabet from $A-E$ denotes tetrad indices on the screen space.

\section{Computational cost in cosmology}\label{sec:GR}

We discuss the cosmological zoom-in N-body simulation within Newtonian gravity in sub-section \ref{sec:zoom-in} and describe focusing singularity(shell crossing singularity) in general relativity/Newtonian gravity in sub-section \ref{sec:clocks_syn}. We then derive an extension of the gravitational focusing theorem in sub-section \ref{sec:matter_horizon}.  The inevitability of affine parameter change is discussed in sub-section \ref{sec:breakdown_equivalence},

\subsection{Spatial hierarchical and piece-wise dynamical time in Newtonian gravity}\label{sec:zoom-in}

Given an ensemble of $N$ gravitationally interacting particles, the gravitational force experienced by each particle $i$ with mass $m_i$ at position $\mathbf{r}_i$ due to  the presence of all other particles $j$ is given by :
\begin{equation}\label{eq:force_eqn}
\mathbf{F}_i = -G m_i \sum_{j \neq i}^N \frac{m_j (\mathbf{x}_i - \mathbf{x}_j)}{|\mathbf{x}_i - \mathbf{x}_j|^3}\,,
\end{equation}
where $G$ is the gravitational constant. The naive computational cost of calculating $\mathbf{F}_i$ for all $N$ particles requires $O(N^2)$ operations, as each particle interacts with $N-1$ others. In the limit of large $N$ (say  $10^{23} $),  the operation is computationally expensive even at a single time slice. 
It was discovered in the 1980s that by organising the particles into a multi-level, nested structure and treating distant particles as a group and approximating their collective gravitational force using their centre of mass and total mass, the computational cost reduces to $O(N \log N)$ \cite{1986Natur.324..446B}.
Once the acceleration ${\bf{a}}_i = {\bf{F}}_{i}/m_i =  \d^2 {\bf{x}}_i /\d t^2 /m_i = \d {\bf{v}}_{i}/\d t/m_i$ (${\bf{v}}$ is the velocity)
 for each particle is known at a given time step, the simulation updates their positions and velocities for the next time step. The computational cost of time integration scales like $\mathcal{O}\left( N\right)$, however, the dynamical timescale is environment-dependent, $t_{\rm{dyn}} \sim 1/\sqrt{G\rho}$, where $\rho$ is the matter density.
Therefore, the total computational cost per timestep in an N-body simulation includes the cost of force calculation and the cost of time integration
\begin{equation}\label{eq:gravitataionl_cost}
{\rm{Force~evaluation~cost}} = {\bf{F}}\left[ N,\<N_t\>  \right]  \sim\mathcal{O}\left(\<N_t\>\times N \log N \right)
\end{equation}
where $\<N_t\> =  {\left[{T_{\rm{final}}}/{\Delta t}\right]}$ is the total number of time steps. $T_{\text{final}}$ is the total physical time to be simulated,  $\Delta t$ is the simulation time step, which typically scales with resolution as $\Delta t \propto \Delta x^{3/2}$ due to gravitational dynamical timescales.

The particles in an N-body simulation experience a wide range of dynamical timescales. In a dense environment like a galaxy cluster, dynamical timescales are shorter when compared to particles in a low-density environment~\cite{Navarro:1994MNRAS.267..401N}. The diversity in dynamical timescales(proportional to resolution) leads to a fundamental computational tension in cosmological N-body simulations. To resolve an over-dense region within a large volume simulation box $V \sim L^3$ ($L$ is the length), require small $\Delta x$ thereby, increasing the number of particles $N  = \left( {L}/{\Delta x} \right)^3$ .  Increasing the number of particles increases the memory requirement needed to store positions, velocities, and the overall computational cost \cite{Klypin:1997sk,Springel:2005mi,Dehnen:2000nh}.
 More particles allow finer resolution but increase computational demands; hence, there exists a fundamental computational tension between covering a large volume and high resolution.

The critical driving force for computational resources is the diversity in dynamical timescales.  There are two main classes of time step integrators used to update the particle position: Fixed time step integrators and Adaptive time step integrators~\cite{Zemp:2006vh}.
We focus on fixed-time step integrators because they preserve the time reversal symmetry ($t\to -t'$) of the Newtonian force equation (equation \eqref{eq:force_eqn}) being solved. They also preserve the simplectic structure of the Hamiltonian and conserve energy over long times.   There are other methods of time step integrators that fall between these two~\cite{1998MNRAS.297.1067K,2024OJAp....7E...1P}. 
A typical example of the fixed-time step integrator is the {Kick--Drift--Kick (KDK)} leapfrog scheme, which proceeds as
\begin{eqnarray}\label{eq:global_timesteps1}
  \mathbf{v}^{n+1/2} &=& \mathbf{v}^n + \frac{\Delta t}{2} \, \mathbf{a}(\mathbf{x}^n)\,,
  \\
    \mathbf{x}^{n+1} &=& \mathbf{x}^n + \Delta t \, \mathbf{v}^{n+1/2}\,,
    \\
      \mathbf{v}^{n+1} &=& \mathbf{v}^{n+1/2} + \frac{\Delta t}{2} \, \mathbf{a}(\mathbf{x}^{n+1})\,,
      \label{eq:global_timesteps3}
\end{eqnarray}
where  the forward oreientation is always chosen $\Delta t >0$ for $t: -\infty \to \infty$. The choice of the time step is made at the initial time. Since $\Delta t \propto \Delta x^{3/2}$, there exists a time in the future $t_{\star}$ for some local regions $r < r_{\star}$ where the global step will cease to be optimal.   This usually manifests as turnaround time for a class of particles, that is, when particles within an overdense region lose their synchronisation to the global clock or fail to catch up with the Hubble flow.

The cosmological zoom-in N‑body simulation was developed as an optimal solution for exploring over-dense regions at high resolution with fewer computational resources~\cite{Katz:1994MNRAS.270L..71K,Navarro:1994MNRAS.267..401N}.
The strategy is to simulate a large volume (the universe) at low resolution first.
Then, a region of interest (e.g. a cluster) is identified within the low-resolution simulation and re-simulated at high resolution. 
The process involves tracing the particles back to the initial conditions and then evolving them forward in time at a higher resolution as a separate universe with a background spacetime defined by the low-resolution simulation~\cite{Hahn:2011uy}.  
\begin{figure}
\includegraphics[width=100mm,height=70mm] {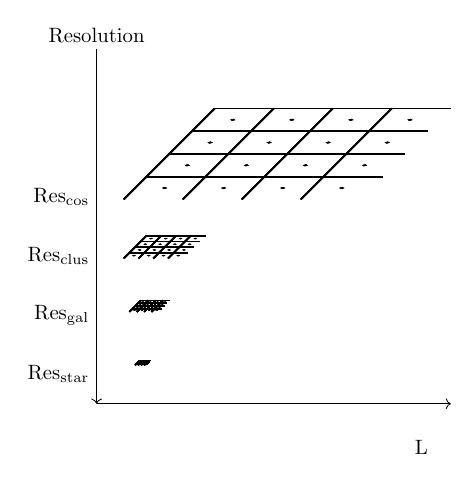}
\caption{A sketch of the tractable resolution as a function of scale for zoom-in simulation. A large volume is simulated at resolution ${\rm{Res}}_{\rm{cos}}$ determined by the mass element. The dynamics of the sub-structure is probed by dividing the grid into sub-grids and sampling at smaller mass elements. The procedure is sequenced.  }
\label{fig:resolution}
\end{figure}
 The intractable computational cost  of getting a large volume at high resolution simulation is reduced substantially by the optimal sequencing of the number of particles in nested layers(see Figure \ref{fig:resolution})
\begin{equation}\label{eq:zoom-in-cost}
\underbrace{N\left[\frac{L_{\text{low}}}{\Delta x_{\text{low}}}\right]}_{\text{tractable}} + \cdots +\underbrace{N\left[\frac{L_{\text{high}}}{\Delta x_{\text{high}}}\right]}_{\text{tractable}} \ll \underbrace{N\left[\frac{L_{\text{low}}}{\Delta x_{\text{high}}}\right]}_{\text{prohibitive cost}} \,.
\end{equation}
To appreciate the amount of cost reduction achiavable with the zoom-in approach, we estimate the ratio  of the total number of particles that the cosmological zoom-in N-body simulation requires to achieve a high resolution $\Delta x_{\rm{high}}$ to the total number of uniform resolution large volume simulation box would require:
\begin{eqnarray}
\frac{N_{\rm{Total}} }{N_{\rm{Uniform}}} &=&
 \frac{M_{\rm{Low}} }{M_{\rm{Uniform}}}
\left[  \frac{1}{k_{m}} +  \frac{1}{ k_{V}  k_{\rho}  }\right] \approx  \frac{1}{k_{m}} = \frac{m_{\rm{High}}}{m_{\rm{Low}}}\,,
\end{eqnarray}
 where ${M_{\rm{Low}}} = \rho_{\rm{Low}}V_{\rm{Low}}$ is the total  mass in the low resolution simulation volume $V_{\rm{Low}}$ and $ \rho_{\rm{Low}}$ is the mean mass density. Similarly,  $M_{\rm{Uniform}} = \rho_{\rm{Uniform}} V_{\rm{Uniform}} $ is the total mass within the entire simulation box sampled at high resolution $\Delta x_{\rm{High}}$ and $\rho_{\rm{Uniform}} $ is the corresponding matter density.  We have defined the ratios of some quantity: matter density $k_{\rho}  =\rho_{\rm{Low}}/ \rho_{\rm{RoI}}$,  volume $k_{V} = V_{\rm{Low}}/V_{\rm{RoI}}$ and mass element  $k_{m} = m_{\rm{Low}}/m_{\rm{High}}$. The most important ratio is the ratio of mass elements in the low-resolution box to the mass elements in high resolution box. In a typical low-resolution N-body simulation, a typical region of interest is in the range of the size of a typical cluster. The mass element, and the size of a typical cluster is about  $m_{\rm{Low}} \sim10^{12} M_{\odot}$, then to study the internal dynamics of the region of interest such that the dynamics of sub-structures the size of a typcial galaxy with mass of about $\sim 10^{11} M_{\odot}$, the simulation mass element of about $m_{\rm{High}} \sim10^{10} M_{\odot}$ is needed, hence, it reduces the particle count to about $\sim1\%$ of the uniform case.
Similarly, a different and suitable time integration step is used to evolve particles within the zoomed-in region of interest
\begin{eqnarray}\label{eq:fast_timescales1}
  \mathbf{v}^{n+1/2}_{\rm{RoI}} &=& \mathbf{v}^n_{\rm{RoI}} + \frac{\Delta t_{\rm{RoI}}}{2} \, \mathbf{a}_{\rm{RoI}}(\mathbf{x}^n_{\rm{RoI}})\,,
  \\
    \mathbf{x}^{n+1}_{\rm{RoI}} &=& \mathbf{x}^n_{\rm{RoI}} + \Delta t_{\rm{RoI}} \, \mathbf{v}^{n+1/2}_{\rm{RoI}}\,,
    \\
      \mathbf{v}^{n+1}_{\rm{RoI}} &=& \mathbf{v}^{n+1/2}_{\rm{RoI}} + \frac{\Delta t_{\rm{RoI}}}{2} \, \mathbf{a}_{\rm{RoI}}(\mathbf{x}^{n+1}_{\rm{RoI}})\,,
      \label{eq:fast_timescales3}
\end{eqnarray}

While the cosmological zoom-in N-body approach offers a powerful strategy for reducing computational costs, its underlying physics has been limited to Newtonian gravity. Within this framework,  particles are treated as simple point masses, inherently lacking the internal degrees of freedom necessary to describe the full complexity of their interactions.
A more comprehensive and physically robust understanding emerges in GR. In GR, particles follow timelike geodesics in spacetime, and GR  predicts that a particle may cease to follow a geodesic within a finite proper time. This breakdown is mathematically connected to the formation of the matter horizon, a causal boundary beyond which the trajectory of a particle with a given initial data can no longer be trusted as a geodesic. The emergence of a shell crossing singularity in GR is a direct consequence of extending a one-parameter family of geodesics beyond this matter horizon~\cite{Ellis:2010fr,Umeh:2023lbc}.

The concept of horizons, however, is not exclusive to GR. Precursors can be found in Newtonian gravity, where they manifest as boundary conditions. For instance, the notion of escape velocity exceeding the speed of light served as a conceptual forerunner to the event horizon of a black hole~\cite{HawkingandEllis:1973lsss.book}. Similarly, the matter horizon in GR finds its analogue in Newtonian gravity, also in the form of escape velocity determined by the total mass of the system $v_{\rm{esc}} = \sqrt{2GM/R}$, where $M$ is the total mass and $R$ is the radius. This is locally determined; test particles with velocity less than $v_{\rm{esc}} $ are trapped.
The fundamental shortcoming of Newtonian gravity is its inability to provide a self-contained mathematical structure for extending the spacetime beyond this boundary. While the cosmological zoom-in N-body approach offers a heuristic argument for probing small scales by isolating and managing timescales for gravitational interaction, it lacks a rigorous theoretical foundation.
This paper's central objective is to demonstrate that the existence of a matter horizon in general relativity provides the precise mathematical grounding underpinning the cosmological zoom-in approaches. We will show that the strategy for avoiding gravitational focusing singularities in GR is mathematically equivalent to the heuristic strategy adopted in cosmological zoom-in N-body simulations to mitigate computational costs.

\subsection{Point-particle approximation in general relativity and focusing singulairty}\label{sec:clocks_syn}

The general relativistic cosmological N-body simulations that describe structure formation in the universe are based on the standard Einstein-Hilbert action  coupled to an ensemble of particles~\cite{Adamek:2014xba,Adamek:2016zes}
\begin{eqnarray}\label{eq:GR_eqns}
R_{ab} - {1 \over 2} g_{ab} R+\Lambda g_{ab}= \kappa T_{ab}\,,
\end{eqnarray}
where $\kappa = 8\pi G/c^4$ is the gravitational coupling constant,   $R_{ab}$ is the Ricci tensor, $R$ is the Ricci sccalar and  $T_{ab}$ is the energy-momentum tensor.
The biggest task in cosmology is how to model the matter content of the universe, i.e. $T_{ab}$.  The nature and composition of $T_{ab}$ determine the local curvature and geometry of the universe. The most common approach is to assume that the matter content of the universe is constructed from an ensemble of point  particles $S_{\text{matter}} =\sum_{\ell} S_{\ell} $, 
where $S_{\ell}(x^a_{\ell},{x^a_{\ell}}')$  is the  action for the $\ell$-th massive particle ~\cite{Adamek:2016zes}
\begin{eqnarray}\label{eq:massive_particle_actionl}
S_{\ell}(x^a_{\ell},{x^a_{\ell}}')  = -m_{\ell} \int_{\tau_i}^{\tau_f}  \sqrt{-g_{ab} \frac{\d x^a_{\ell} }{\d\tau_{\ell}}\frac{\d x^b_{\ell}}{\d\tau_{\ell}} } \d\tau_{\ell} \,,
\end{eqnarray}
where   $\tau_{i}$  and $\tau_{f}$ are the initial and final proper time of the particle, ${u^a_{\ell}} = {\d x^a_{\ell}}/{\d\tau_{\ell}}$, 
$x_{\ell}^a(\tau_{\ell})$ is the spacetime trajectory of the massive $\ell$-th particle, $m_{\ell }$ is the rest mass of the ${\ell}$-th particle and $\tau_{\ell}$ is the proper time for the ${\ell}$-th particle.  
 Note $x^a$ is a point in the spacetime and the delta function is non-zero only when $x^a$ coincides with  $x^a_{\ell}(\tau_{\ell})$: 
The  total energy-momentum tensor is constructed from the action of a massive particle equation \eqref{eq:massive_particle_actionl}) :  $ T^{ab}  =  \sum^N_{\ell} T^{ab}_{\ell} $, where 
 \begin{eqnarray}\label{eq:EMT_particles}
 T^{ab}_{\ell}  
 \equiv-\frac{2}{\sqrt{-g}}
  \frac{\delta\left(\sqrt{-q} {L}_{M\ell}\right)}{\delta g_{ab}} 
 =  { m_{\ell}}  \int_{\tau_i}^{\tau_f} \d \tau
 \frac{\d x^a }{\d\tau}(\tau)
 \frac{\d x^b}{\d\tau} (\tau)
\frac{\delta^{(4)}\left(x^i - x^i_{\ell}(\tau)\right) }{{\sqrt{-g(t,x^i)}}}\,, 
\end{eqnarray}
where ${\sqrt{-g} }$ is the is square root of the determinant of the spacetime metric tensor,  $\delta^{(4)}\left(x^a - x^a_{\ell}(\tau_{\ell})\right)$ is the 4-D Dirac delta function, it is normalised to unity.  
The equation of motion resulting from equation \eqref{eq:massive_particle_actionl}  is the basis of the Newtonian N-body simulation~\cite{Springel:2005mi,2011EPJP..126...55D}.
The massive particle action given in equation \eqref{eq:massive_particle_actionl} is used in cosmology to propagate particles or fluid mass elements. In the point particle limit, the finite extent is neglected, and the particle is treated as a test particle on a given background spacetime.

In the remainder of this subsection, we describe how this approximation breaks down on scales comparable to the physical size of a gravitationally bound system.  We shall also describe various approaches that have been developed over the years to address this limitation. 
We start with the action of a massive particle given in equation \eqref{eq:massive_particle_actionl}. The first infinitesimal variation gives the geodesic equation~\cite{Umeh:2023lbc}.
\begin{eqnarray}\label{eq:geodeisc_eqn}
u^d\nabla_d u^c = 0 
\,.
\end{eqnarray}
where we imposed proper variation at the endpoints.  The massive particle trajectory is geodesic within $[\tau_{\rm{ini}},\tau_f]$ provided equation \eqref{eq:geodeisc_eqn} holds.  The affine parameter, $\tau $  is the proper time measured by an observer at rest with the particle. 
 Equation \eqref{eq:geodeisc_eqn}  is unchanged under the reparameterisation of the affine parameter. 
The reparameterisation invariance induces a scale transformation of the velocity vector 
$\tau = W\tilde{ \tau} + Z\,,
 u^a =  \frac{1}{W} \tilde{u}^a \,,$
 where $W$ and $Z$ are non-zero free parameters.  
 Equation \eqref{eq:geodeisc_eqn} is a set of second-order differential equations; therefore, in addition to the freedom to fix $W$ and $Z$,  there is a freedom to choose the initial position, $x^a( \tau_{\rm{ini}})$ and the initial velocity, $ u^a_{\rm{ini}}$.  These coordinate choices are central to our discussion. 
The geodesics equation (i.e. equation \eqref{eq:geodeisc_eqn}) reduces to the  equations of motion for the Newtonian N-body simulation, which we used to define the force evaluation cost function in equation \eqref{eq:gravitataionl_cost}
\begin{eqnarray}\label{eq:Nbody}
\frac{ \d {\bf{x}}_{i} }{\d t} &=& \frac{{\bf{p}}_{i}}{m_{i}}\,,
\qquad \qquad 
\frac{\d {\bf{p}}_{i}}{\d t}  = - m_i{\bf{ \nabla}} \Phi ({\x}_{i}) \,, 
\qquad {\rm{with}}\qquad 
\Phi(x_{i}) = - G \sum_{ j = 1}^{N} \frac{m_{j}} { \left[ \left(x _{i}- x_j\right)^2 \right] }\,,%
\end{eqnarray}
where  $ \Gamma^{a}{}_{bc} u^b u^c = -\delta^{ab}(\partial_{b} g_{00})/2$ and $g_{00} = -(1+2 \Phi)$.
The initial conditions for equation \eqref{eq:Nbody} are set such that the simulation starts with a specific initial distribution of particles determined by the early universe physics using the Boltzmann solver~\cite{Lewis:1999bs,Lesgourgues:2011rg}. 
Then equation \eqref{eq:Nbody} gives the current position of the particle $x^i$ in terms of the initial position
\begin{eqnarray}\label{eq:displacement_field}
x^i(t,q^i) = q^i + \Psi^i(t,q^i)\,,
\end{eqnarray}
where $q^i$ is the initial position of the particle, which is assumed to be an FLRW spacetime,  $\Psi^i$ is known as the displacement field.  
Imposing mass conservation $\rho(t, x^i) \d^3 x =\bar{\rho}(t)\d^3 q$ and assuming that a collection of discrete elements (Lagrangian fluid elements or particles) is moving under gravity in an expanding background immediately provides a link between matter density at the current time $\rho(t, x^i) $  and the initial matter density $\bar{\rho}$. 
\begin{eqnarray}\label{eq:mass_conservation}
1+ \delta(t,{\x}) = \bigg|\frac{\d^3 x}{\d^3 q}\bigg| = \frac{1}{{{\rm{det}}[\mathcal{J}](t,{q}) }}\,,
\end{eqnarray}
where $\delta \equiv \delta \rho/ \bar{\rho}$ is the  matter density contrast or the fluctuation in the mass density with respect to the mean density $\bar{\rho}$, 
${{{\rm{det}}[\mathcal{J}](t,{q}) }} = \big|{\d^3 q}/{\d^3 x}\big| = {\rm{det} \left[ \delta_{ij}+ {\partial \Psi_i}/{\partial q_{j}}\right]} $ is the determinant of the Jacobian of transformation from initial coordinates $q^i$ to the coordinates at the current time $x^i$.  The key assumption within standard cosmology is that at early times, the overdensity was very small, i.e. the spacetime was almost FLRW.  The overdensity then grew with time due to gravitational  instability~\cite{Zeldovich:1969sb,Takada:1999cg}.   
The groth of density contrast tends to infinity $\delta(t,{\x}) \to \infty$ as ${{{\rm{det}}[\mathcal{J}](t,{q}) }} \to 0$ at finite time(this is what we refer to as the gravitational focusing singularity or caustics).

The shell crossing manifests in the nonlinear regime of structure formation, and various strategies are used to regulate the singularity, for example, 
\begin{itemize}
\item  {\tt{Standard  Perturbation Theory (SPT)}}: This approach expands the equations and solves the leading multipole moment(neglects anisotropic stress tensor) of Vlasov equation in powers of $\delta$~\cite{Bernardeau:2001qr}. However, when mode coupling becomes important on small scales ($k \ge 0.1$[h/Mpc]), a momentum cut-off is used to suppress diverging contributions from small-scale modes~\cite{Umeh:2015gza,Umeh:2016thy,Umeh:2020zhp,Umeh:2021xqm}.  

\item {\tt{The Effective Field Theory of Large-Scale Structure  (EFTofLSS)}}:  It provides a more systematic way of accounting for the impact of short-scale physics on long-scale observables.
It includes a parametrised anisotropic stress tensor and interprets parameters as capturing unknown short-scale physics on quasi-linear scales~\cite{Perko:2016puo}. EFTofLSS introduces a large number of parameters at higher loop order in redshift space, and a smoothing scale is usually fixed at about the quasilinear scales~\cite{Pajer:2013jj}.  
 \item {\tt{Vlasov perturbation theory (VPT)}}:  VPT addresses the shortcomings of SPT by going beyond fluid approximation, by capturing the contribution of higher cumulants (or moments) of the phase-space distribution function. 
 ~\cite{Garny:2022tlk,Garny:2022kbk,Garny:2025zlq}.
VPT offers a more analytical route to understanding small-scale structure formation in the Newtonian limit. It supposes the existence of a scale below which the background dynamics differ completely from the global FLRW spacetime due to non-vanishing vorticity contribution and backreaction. 
 
\item {\tt{Cosmological N-body simulation}}: Similar to the discussion in sub-section \ref{sec:zoom-in}, the cosmological N-body simulation solves equation \eqref{eq:Nbody}. The shell crossing singularities occur during close encounters ($r \to 0)$, where the force
 $F \propto {1}/{r^2}$ diverges. The gravitational softening length ($\varepsilon$
) is used to smooth out the potential \cite{Dehnen:2000nh}. 
This parameter modifies the Newtonian gravitational potential(equation \eqref{eq:Nbody}) at scales below the softening length, effectively regularising the force calculation. While a larger softening length enhances numerical stability by suppressing artificial singularities, it imposes a minimum resolvable mass scale, limiting the simulation's mass resolution \cite{Zhang:2018nqh}. Standard configurations often set the softening length on the order of a few comoving kpc/h \cite{Zhang:2018nqh}.

\item {\tt{ Cosmological zoom-in simulations}}: We provided an extended discussion in sub-section \ref{sec:zoom-in} on how the cosmological zoom-in N-body simulations probe small scales by re-simulating a region of interest at higher resolution~\cite{Wetzel:2022man}. It employs coarse softening within a large-volume, low-resolution background, subsequently re-simulating a region of interest (e.g., a forming halo) with higher resolution and finer softening length.
\end{itemize}

Finally,  equation \eqref{eq:mass_conservation} is derived within the framework of Newtonian gravity and assumes a flat spatial geometry at early time and smooth evolution. It does not account for the curvature of spacetime or the local relativistic effects described by GR. The spacetime curvature only enters through the FLRW scale factor, not through the local dynamics.
Secondly, the assumption that the mapping $q^i \mapsto x^i$ is smooth and invertible until shell crossing (when ${\rm det}[\mathcal{J}] \to 0$) also does not account for the general relativity effects, such as the matter horizon. This is essentially the point we plan to establish: that the matter horizon forms before the shell crossing singularity and that GR provides more mathematical tools to avoid the shell crossing singularity.

\subsection{Matter horizon  in general relativity}\label{sec:matter_horizon}

For a single particle in the early universe, say a Hydrogen atom, the backreaction of the particle propagation on the spacetime is negligible; however, when $10^{60}$ Hydrogen atoms are bound together by gravity, their impact on the spacetime becomes non-negligible on the scale proportional to the size of the gravitationally bound system. One way to study this is by performing a second variation(the rate of change of nearby geodesics).  It carries the information about whether the one-parameter family of geodesics are contracting or expanding. This is a critical feature that determines whether there will be a gravitational focusing singularity or not~\cite{HawkingandEllis:1973lsss.book}. 
The critical points of the second variation of equation \eqref{eq:massive_particle_actionl} give the geodesic deviation equation~\cite{Umeh:2023lbc}: 
\begin{eqnarray}
 \frac{\d^2 \xi^{c}}{\d\tau^2} + R^{c}{}_{def} \xi^{d} u^{e} u^{f}  = 0 \,,
\end{eqnarray}
where $ R^{a}{}_{def}$ is the Riemann tensor and $\xi^{d}$ is the deviation vector, it measures how nearby geodesics differ from the central geodesic.  We assumed that the endpoints of the geodesics are held fixed.
Lie dragging $\xi^a$ along the integral curves of $u^a$ gives~\cite{HawkingandEllis:1973lsss.book}
$\mathcal{L}_{u} \xi^a = 0 \rightarrow  u^b\nabla_b \xi^{a}  = \nabla^{b} u^a \xi^b  $.
We then  decompose $ \nabla_{b} u_{a} $  into irreducible coordinate independent physical quantities following~\cite{Ellis:1998ct,Ellis:1990gi}, 
\begin{eqnarray}\label{eq:decomposeCDU}
 \nabla_{b} u_{a} = - u_b A_a + \frac{1}{3} \Theta h_{ab } +\sigma_{ab} \,,
\end{eqnarray}
where ${h}_{ab}$ is the metric on the hypersurface orthogonal to $u^a$, $A^a = u^d\nabla_d u^a$ is the acceleration,  it vanishes by equation \eqref{eq:geodeisc_eqn}.  $\Theta  = \nabla_{a} u^{a} $  is the expansion scalar, it describes the tendency of one-parameter family of geodesics to expand $\Theta >0$  or contract/collapse $\Theta<0$.
 $\sigma_{ab} = h_{a}{}^{c} h_{b}{}^{d} \nabla_{\<c} u_{d\>}$ is the shear deformation tensor, it describes the rate of change of the deformation of a one-parameter family of geodesics when compared to flat spacetime.   We neglected the vorticity tensor for simplicity.  
Expressing the geodesic deviation equation in terms of $\Theta$ and $\sigma_{ab}$ gives the following equations of motion ~\cite{2012reco.book.....E,Ellis1971grc..conf..104E,Ellis2009,Ellis:1998ct}:
\begin{eqnarray}
\frac{ {\rm{D}} {{\Theta}} }{{\rm{D}} \tau} &=& - \frac{1}{3}{\Theta}^2 - {\sigma}_{ab}{\sigma}^{ab} 
- {R}_{ab} {u}^a {u}^b\,,
\label{eq:expansion_equation}
\\
\frac{ {\rm{D} } {{\sigma}}_{ab}}{{\rm{D}} \tau} &=& - \frac{2}{3} {\Theta} {\sigma}_{ab} - {\sigma}^{c}{}_{\<a}{\sigma}_{b\>c} 
 - {C}_{acbd}{u}^c {u}^d\,,
\label{eq:shear_equation}
\end{eqnarray} 
where ${C}_{acbd}$ is the Wely tensor  and $R_{ab}$ is the Ricci tensor.   
Equations  \eqref{eq:expansion_equation} and  \eqref{eq:shear_equation}  are first-order differential equations, to solve them,  we need to specify the initial values of $ \Theta _{\rm{ini}}$ and $  {\sigma}_{{\rm{ini}}ab}
$. However, it is possible to solve equation \eqref{eq:expansion_equation}  for the general values of the initial values provided physical conditions are placed on the energy density $ {R}_{ab} {u}^a {u}^b \ge 0$(weak  energy condition) ~\cite{HawkingandEllis:1973lsss.book,Poisson:2003nc}
\begin{eqnarray}\label{eq:focussing_theorem}
\frac{ {\rm{D}}  \Theta^{-1}}{ {\rm{D}} \tau}  \geq \frac{1}{3} \qquad {\rm{with~a~solution }} \qquad 
 \frac{1}{\Theta } \geq \frac{1}{\Theta _{\rm{ini}}}+ \frac{\tau - \tau_{\rm{ini}} }{3}\,,
\end{eqnarray}
 The contributions of the shear tensor to the propagation equation for $\Theta$ is positive definite $\sigma_{ab}\sigma^{ab} \ge  0$. Equation \eqref{eq:focussing_theorem}
predicts a gravitational focusing singularity at a finite time for a class of geodesics which were initially converging $ 0<\Theta _{\rm{ini}}$.
Using the normalised derivative of the determinant of the Jacobian with respect to  the proper time, we can relate 
$ {\rm{det}}[\bf{\mathcal{J}}]$ to $\Theta$~\cite{Umeh:2023lbc}
\begin{eqnarray}\label{eq:J_theta}
\frac{1}{{\rm{det}[\bf{\mathcal{J}}]}} \frac{\d {\rm{det}[\bf{\mathcal{J}}]}}{\d\tau} = \Theta\,.
\end{eqnarray}
 and equation \eqref{eq:focussing_theorem} , gives
   \begin{eqnarray}\label{eq:censorrship}
{{\rm{det}}[\mathcal{J}](\tau,{\x}) } 
 \le { {\rm{det}}[\mathcal{J}](\tau_{\rm{ini}},{\q}) } \left[1 + \frac{{\Theta_{\rm{ini}}}(\tau - \tau_{\rm{ini}})}{ 3}\right]^3  \,,
\end{eqnarray}
where ${\rm{det}}[\mathcal{J}](\tau_{\rm{ini}},{\q})$ is the Jacobian at the initial time.   For  negative ${ \Theta _{\rm{ini}}}<0$,  we immediately see that ${{\rm{det}}[\mathcal{J}](\tau,{\x}) } \to 0$
  at $\tau  \leq 3/ \Theta _{\rm{ini}}$.  Hence, for negative $\Theta_{\rm{ini}}<0$,  gravitational focusing singularity is inevitable in the future $\tau >\tau_{\rm{ini}}$. This is the gravitational focusing theorem ~\cite{Penrose:1969pc,HawkingandEllis:1973lsss.book}.

 In the context of large-scale structures of the universe, however,  the metric perturbation generated during cosmological inflation provides the seeds for structure formation.  Describing the universe from the inflationary epoch would require modelling quantum fluctuations and their transition to classical perturbations; hence, calculating $\Theta_{\rm{ini}}$ from the first principle would involve quantum gravity and a precise treatment of reheating~\cite{Ellis:2015wdi}. 
 Within the standard cosmology, however, it is consistent to set the initial conditions for the formation of large-scale structures deep in the matter-dominated era when modes have re-entered the horizon~\cite{Bernardeau:2001qr, Springel:2005mi,Bruni:2013qta}. For small-scale modes, however, many of them re-entered during the radiation-dominated era~\cite{Fidler:2015npa}. 
It is shown in~\cite{Fidler:2015npa} that the effect of radiation can be incorporated during post-processing  ~\cite{Fidler:2015npa}.   Therefore, we adopt the well-known Zeldovich approximation for setting up initial conditions for structure formation with the understanding that our conclusion will remain largely unchanged when the impact of the radiation-dominated epoch and inflationary cosmology is included~\cite{Zeldovich:1969sb,Springel:2005mi}.
 Using equation \eqref{eq:displacement_field}, the  initial velocity becomes
\begin{eqnarray}\label{eq:velocity field}
v^i_{\rm{ini}}({\q})  \equiv v^i(t_{\rm{ini}},{\q}) = H(t_{\rm{ini}}) x^i(t_{\rm{ini}},q^i)  + v^i_{\rm{rel}}(t_{\rm{ini}},{\q}) \,,
\end{eqnarray}
where $v^i_{\rm{rel}}(t,{\q})  = H(t)f(t)\Psi^i(t,{\q})$ is the relative velocity between the expanding background spacetime and perturbation.  The displacement vector field is given in terms of the density contrast  $\delta({\k})$: $ \Psi({\k},t_{\rm{ini}}) =  - i {\k} \tilde{\delta}(t_{\rm{ini}},{\k})/{k^2}$. The Fourier transform of the density contrast  $\delta({\k})$ is given by  the matter power spectrum $P(k)$~\cite{Malik:2008im}
$
\tilde{\delta}_{m}({\k}) = \sqrt{P(|{\k}|) } \mu(|{\k}|)\,,
$
where  $P(k)  = \Delta^2 k^{n_s-1} T^2(k)$, where $ \Delta^2$ and $n_s$ are the amplitude and the spectral index of the initial density fluctuation, respectively. These two values are determined by the physics of the early universe.  $T(k)$ is the transfer function, and $\mu(r)$ )  is white noise sampled from a Gaussian distribution with zero mean and unit variance. The real space over density is obtained by performing an inverse Fourier transform of $\tilde{\delta}_{m}({\k}) $: $\tilde{\delta}_{m}({\r})  = \mathcal{T}({\r}) \star\mu({\r})$.

 The important point to note is that  for a universe similar to ours, $\Theta$ is composed of two parts: 
\begin{equation}\label{eq:decomp_Theta}
  \Theta = \Theta_{H}+ \Theta_{L}
\end{equation}
where $\Theta_H$ denotes the global part, i.e $\Theta_H = 3 H$(it is determined by chossing foward time FLRW  background FLRW spacetime, this is a coordinate choice) and  $\Theta_{L}$  denotes the local expansion.
At the initial hypersurface, $\Theta_{L} = {\partial}_i v^{i} _{\rm{ini}}$, where $v^a$ is the relative velocity between the global spacetime and the local matter distribution~\cite {Ellis:2002tq}.
In general, $\Theta_{\rm{ini}}$ is always positive for a universe with coordinate time flowing forward
\begin{eqnarray}\label{eq:initial_theta}
\Theta_{\rm{ini}} \approx 3 H_{\rm{ini}} +{\partial}_i v^{i} _{\rm{ini}} >0 \,.
\end{eqnarray}
where  $H_{\rm{ini}} = H(\tau_{\rm{ini}})$  is the Hubble rate at the initial time and $v^{i} _{\rm{ini}} = v^i(\tau_{\rm{ini}},{\q})$ is the initial peculiar velocity field. 
\begin{figure}[h]
\centering
\includegraphics[width=100mm,height=50mm] {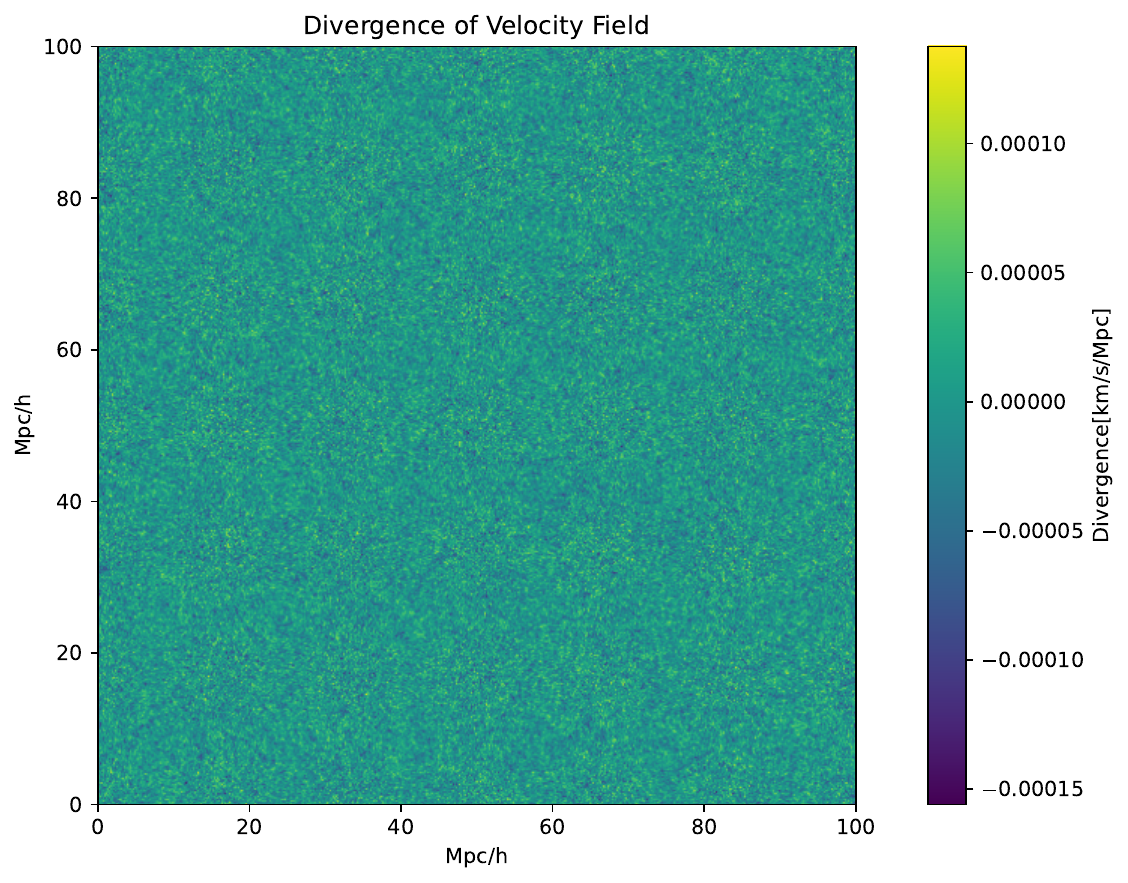}
\caption{
We show the divergence of the initial relative velocity vector field(i.e equation \eqref{eq:velocity field}) $\Theta_{\rm{ini L}} ={\partial}_i v^{i} _{\rm{ini}} $..
Note that $\Theta_{\rm{ini L}}$ has both negative and positive values, and the amplitude is much less than one at the initial hypersurface. 
The total expansion is positive, and it is dominated by the background component.  We evaluated the initial data at a redshift of $z =10$.  
} 
\label{fig:initial_conditions}
\end{figure}
We show the distribution of the initial values of $ \Theta_{\rm{ini}L} $  in Figure \ref{fig:initial_conditions}.  It is clear that $ \Theta_{\rm{ini}L} $ can be negative and positive.  The negative and positive values denote two possible initial values that will lead to two sub-regions of spacetime in the future.  Sub-regions with $ \Theta_{\rm{ini}L} <0$ are collapsing/contracting regions even the background is expanding, while sub-regions with  $ \Theta_{\rm{ini}L} >0$ are expanding alongside the background spacetime \cite{Ellis:2002tq}.  
The key point here is the global part of $\Theta$ is always poisitive $\Theta_{H}(\tau_{\rm{ini}}) > 0$.  This is due to our choice of a forward-oriented coordinate system.  More discussion of this is presented in section \ref{sec:FLRW_discrete_symmetry}.

To appreciate the physical consequences of both phases of initial values of $ \Theta_{\rm{ini}}$: $ \Theta_{\rm{ini}L} <0$ and $ \Theta_{\rm{ini}L} >0$, we can decompose equation \eqref{eq:expansion_equation} into global and local parts using equation \eqref{eq:decomp_Theta}. The global expansion ${{\Theta_{H}}}$ obey the following propagation equation 
\begin{eqnarray}\label{eq:Global_Raychaudhuri_eqn}
\frac{ {\rm{D}} {{\Theta_{H}}} }{{\rm{D}} \tau} &=& - \frac{1}{3}{\Theta}^2_{H} 
-  \frac{1}{2}\kappa\bar{\rho} - \Lambda \,,
\end{eqnarray}
where ${\sigma}_{ab}{\sigma}^{ab} =0$ has no global part.  We made use of the time-time component of the  Einstein field equation, to express $  u^b u^d R_{bd}$ in terms of the matter density field, $  u^b u^d R_{bd} \approx\frac{1}{2}\kappa \rho + \Lambda  = \frac{1}{2}\kappa\left[\bar{\rho} +  \delta \rho \right] + \Lambda, $ where we have decomposed $ \rho$ into global and local parts $ \rho  = \bar{\rho} + \delta \rho$.  Here $\bar{\rho} $ is the global component given by the mean density of the universe, and $\delta \rho$ is the local part or the fluctuation.
Equation \eqref{eq:Global_Raychaudhuri_eqn} can be solved for $ \frac{1}{2}\kappa^2\bar{\rho} + \Lambda >0$ with the following solution
\begin{eqnarray}\label{eq:Global_focussing_theorem}
\frac{ {\rm{D}}  \Theta^{-1}_{H}}{ {\rm{D}} \tau}  \geq \frac{1}{3} \qquad {\rm{with~a~solution }} \qquad 
 \frac{1}{\Theta_{H} } \geq \frac{1}{\Theta _{H\rm{ini}}}+ \frac{\tau - \tau_{\rm{ini}} }{3}\,,
\end{eqnarray}
where $\tau_{\rm{ini}}$ is the proper time at the initial hypersurface and ${\Theta _{H\rm{ini}}} $ is the initial global expansion.   $1/\Theta_{H}$ vanishes at the finite time $\tau = \tau_{\rm{ini}}  - 3 /\Theta_{H\rm{ini}}  $ only if the initial expansion is negative ${\Theta _{H\rm{ini}}} <0$.  However, the standard cosmology is based on the observation by Edwin Hubble that galaxies recede at speeds proportional to their distance, and it is considered direct evidence that the observed universe is expanding~\cite{Hubble:1929ig}, hence only the ${\Theta _{H\rm{ini}}} >0$ branch is considered.

The local component ${\Theta_L} $ satisfies the following propagation equation 
\begin{eqnarray}\label{eq:local-expansion}
\frac{ {\rm{D}} {{\Theta_L}} }{{\rm{D}} \tau} 
& =   &- \frac{1}{3}{\Theta_L}^2-\frac{2}{3} \Theta_{H}\Theta_{L} - {\sigma}_{ab}{\sigma}^{ab} 
-\frac{1}{2}\kappa\left[  \delta \rho \right] \,,
\end{eqnarray}
where we consider a coordinate system where the proper time associated with the timelike geodesic is synchronised to the coordinate time(Lagrangian frame).
Equation \eqref{eq:local-expansion} includes a term that couples the local dynamics to the global part $ \Theta_{H}\Theta_{L}$. In general,  equation \eqref{eq:local-expansion} is a Riccati differential equation. 
There are different ways to solve it for a given background spacetime; however, it is important to get some insights in the no-coupling limit, i.e $ \Theta_{H}\Theta_{L} = 0$, where equation \eqref{eq:local-expansion} reduces to 
${ {\rm{D}} {{\Theta_L}} }/{{\rm{D}} \tau}  \approx     - {\Theta_L}^2 /3- {\sigma}_{ab}{\sigma}^{ab} 
-\kappa\left[  \delta \rho \right] /2\,.$
hence, it has the same form as the Raychaudhuri equation (equation \eqref{eq:expansion_equation}) with $  u^b u^d R_{bd} $ replaced with $\kappa\left[  \delta \rho \right] /2$.  
 In this limit, it  has a similar solution to equation  \eqref{eq:expansion_equation},
where the weak energy condition is replaced with the assumption that the local region is over-dense $\left[  \delta \rho \right]>0$. 
 \begin{figure}[h]
\includegraphics[width=80mm,height=50mm]{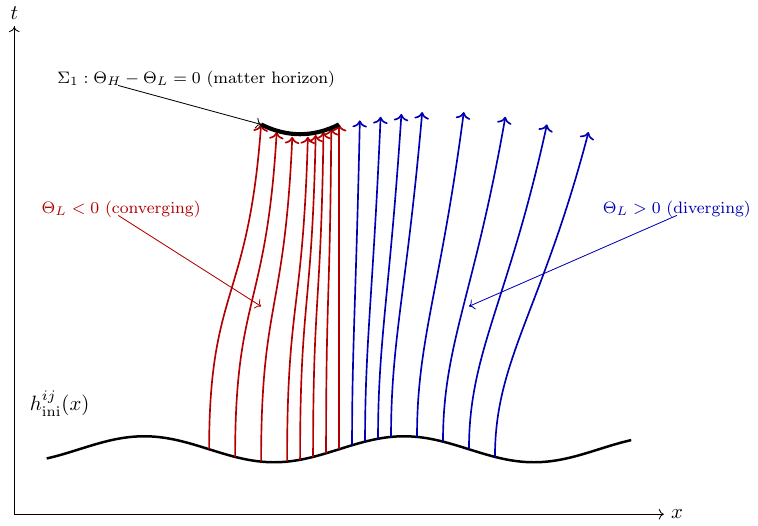}
\caption{
A schematic illustration of the propagation of two families (blue and red) of geodesics orthogonal to a deformed initial hypersurface $h^{ab}_{\rm{ini}}(x)$. The blue family of geodesics emerges from an initially underdense region and expands into the future ($\Theta_{L} >0$.), while the red family of geodesics emerges from an initially overdense region and converges in the future($\Theta_{L} <0$).  
}
\label{fig:converging_diverging}
\end{figure}
Now going beyond the no-coupling approximation, equation \eqref{eq:local-expansion} can be solved similarly to the standard focusing theorem as a partial differential inequality 
\begin{eqnarray}\label{eq:local-expansion3}
\frac{ {\rm{D}} {{\Theta_L}} }{{\rm{D}} \tau} 
+ \frac{1}{3}{\Theta^2_L}+\frac{2}{3} \Theta_{H}\Theta_{L}& \le   & 0 \,,
\end{eqnarray}
where we assume that $\delta \rho >0$.  Again ${\sigma}_{ab}{\sigma}^{ab} >0$ is positive definite.  Integrating equation. \eqref{eq:local-expansion3} leads to 
\begin{eqnarray}\label{eq:focusing_theorem2}
\frac{1}{\Theta_{L}(\tau)} \ge - \frac{1}{\exp\left[\mathcal{I}_{1} (\tau)\right]}  \bigg[ \frac{-1}{\Theta_{L\rm{ini}} }
+\mathcal{I}_{2} (\tau) \bigg]\,,
\end{eqnarray}
where $\mathcal{I}_{1} (\tau) $ is a function of the background expansion
\begin{eqnarray}
\mathcal{I}_{1} (\tau) &=& -\frac{2}{3}\int_{0}^{\tau} \Theta_{H}(\tau') \d\tau'
 \approx   -2 \ln (1+z)
 \,,
\\
\mathcal{I}_{2} (\tau) &=&-\frac{1}{3}\int_{0}^{\tau} \exp\left[\mathcal{I}_{1} (\tau')  \right] \d \tau' 
\approx
-\frac{1}{3}\int_{z}^{\infty}\frac{ \d z}{(1+z)^3 H(z)}    \qquad{\rm{for}}\qquad \tau \in \left[0,\tau_{\rm{max}}\right] \,
\label{eq:I2}.
\end{eqnarray}
In the second approximation, we made use of $\Theta_{H} = 3 H$ and converted proper time $\tau$ to redshift  using  $\d\tau  = -  d z/(1+z)/H(z)$.
The exponential of any real number gives a positive real number; hence, the common factor in equation \eqref{eq:focusing_theorem2} is positive.   For $\Theta_{L\rm{ini}} <0$ that is, for geodesics of particles which started off in the region of negative local curvature of the long wavelength curvature perturbation will appear to converge(become more negative) in finite time given by the condition $\mathcal{I}_{2} (\tau) = 1/ {\Theta_{\rm{ini}} }$.

\begin{figure}[h]
\centering
\includegraphics[width=70mm,height=50mm] { 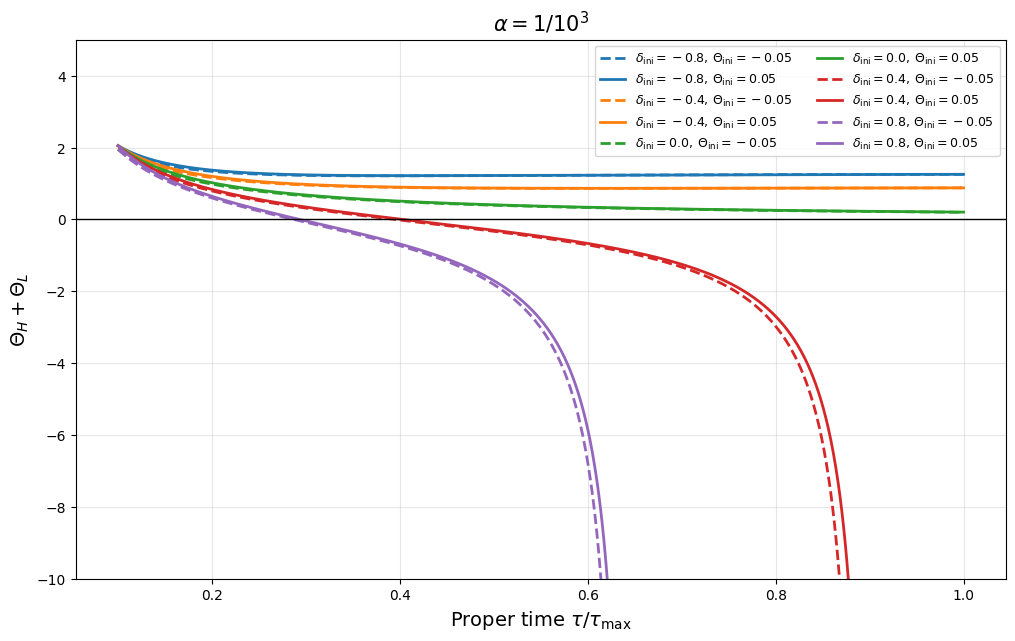}
\includegraphics[width=70mm,height=50mm]{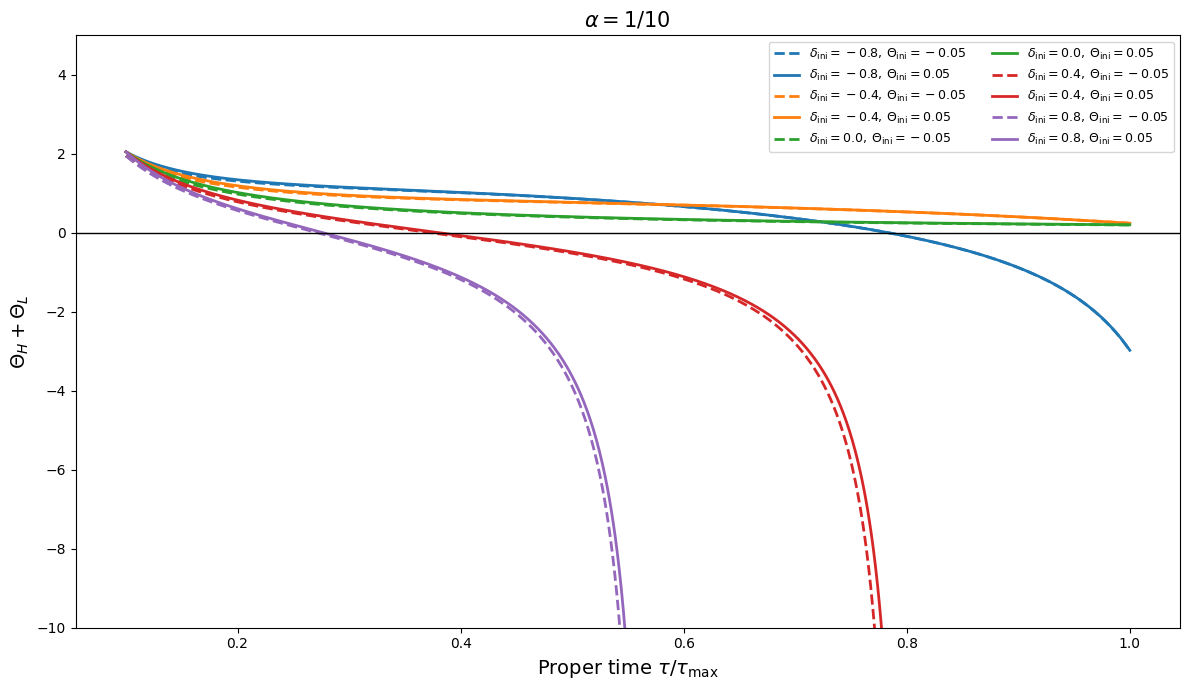}
\includegraphics[width=70mm,height=50mm]{ 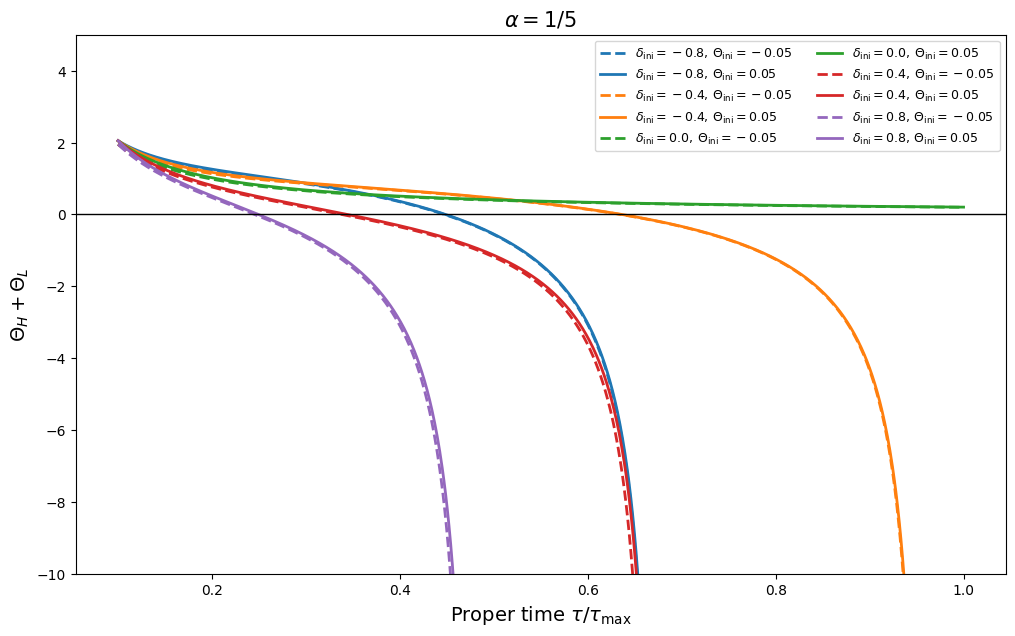}
\includegraphics[width=70mm,height=50mm]{ 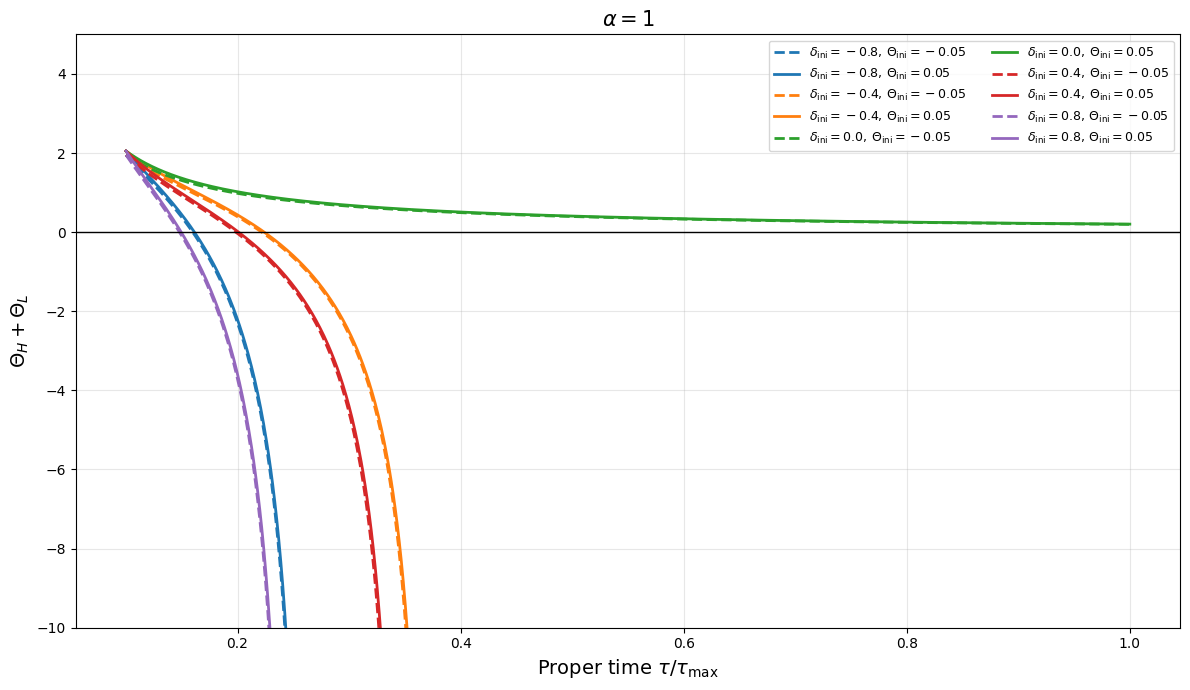}
\caption{The figure shows the plot of the expansion scalar as a function of proper time for various values of the density contrast. The decoupling timescale from the Hubble flow, $\Theta_{H}$, is very sensitive to the initial density contrast. 
} 
\label{fig:matter_horizion}
\end{figure}

Most importantly, we can solve equation \eqref{eq:local-expansion},  in the matter-dominated era under some mild assumptions on the nature of the shear scalar contribution. 
In the matter-dominated era, the Hubble rate is given by $\Theta_{H} = 3 H = 2/\tau$ and the matter density contrast is given by $\delta_{m} = \delta\rho/\bar{\rho} = \delta_{\rm{ini}}\left( {\tau}/{\tau_{\rm{ini}}}\right)^2$, where $\delta_{\rm{ini}}$ is the initial density contrast. 
The evolution of the shear tensor is given by equation \eqref{eq:shear_equation}, however, what contributes to equation \eqref{eq:local-expansion}, is the  shear scalar  and without loss of generallity we can approximate it with ${\sigma}_{ab}{\sigma}^{ab}  =2\alpha \delta_m^2 \Theta^2_{H}/3$, where $\alpha$ is a parameter that regulates the amplitude of the shear scale contribution. In this limit   equation \eqref{eq:local-expansion} reduces to 
\begin{equation}\label{eq:simplified_ODE}
\frac{ {\rm{D}} {{\Theta_L}} }{{\rm{D}} \tau} 
 =   - \frac{1}{3}{\Theta_L}^2-\frac{2}{3} \Theta_{H}\Theta_{L} - \frac{2}{3}\left(\alpha\delta_{m} \Theta_H\right)^2
-\frac{1}{6}\left({ \Theta^2_{H}} \right)  \delta_{m}  \,,
\end{equation}
We made use of the Hamilitonian constraint(Friedmann equation): to  relate  the average density to $\Theta_{H}$:  ${ \Theta^2_{H}} =3 \kappa \bar{\rho} $.   
Given given initial values of $\Theta$ and $\delta_{m}$, we solve equation  \eqref{eq:simplified_ODE} numerical and the  plot of $\Theta$ as a function of time is given in figure \ref{fig:matter_horizion} for different values of $\Theta_{\rm{ini}}$ and $\delta_{\rm{ini}}$. 
The timescale of decoupling from the Hubble flow or the timescale for the formation of the matter horizon is mainly determined by the matter density contrast and the rate of tidal deformation. 
In the spherically symmetric limit, i.e $\sigma_{ab}\sigma^{ab}\approx 0$, the matter density contrast alone determines when a local region decouples from the Hubble flow. The regions with large initial density contrast decouple first from the Hubble flow. The under-dense regions remain in the Hubble flow.   These changes in the presense of tidal deformation. The underdense region decouples too from the Hubble flow, but at much later times.   The regions with zero density contrast remain in the Hubble flow as expected.

The essence of equation \eqref{eq:focusing_theorem2} is the impact it has on the structure of the local spacetime. 
This can be seen by plugging equation \eqref{eq:focusing_theorem2} in equation \eqref{eq:decomp_Theta}, it becomes immediately obvious that geodesic which started off in a local region with $\Theta_{L\rm{ini}} <0$  in the past will decouple from the global Hubble flow $ \Theta = 0$ at finite time.
In analogy to apparent horizons in black hole physics~\cite{Altas:2021ine},  the vanishing of $\Theta$ at finite time indicates the presence of quasi-local horizons (matter horizon).   Quasi-local horizons are horizons defined based on the local structure of spacetime curvature as determined by integral vector fields on the spacetime. Their property differs from that of global horizons, which are defined for the entire spacetime. 
In sub-section \ref{sec:breakdown_equivalence}, we show that this sub-family of geodesics ($\Theta_{L\rm{ini}} <0$) ceases to be geodesic when extended beyond the matter horizon. Also, we show that it is only when these geodesics are extended beyond the matter horizon that shell crossing becomes inevitable.

\subsection{Matter horizon within standard cosmology}\label{eq:particle_formation}

The expansion scalar, $\Theta$, can also be estimated within standard cosmology.
In this case,  we consider a  perturbed metric tensor in Poisson(conformal Newtonian) gauge ~\cite{Umeh:2010pr}
\begin{eqnarray}\label{eq:metric}
\d  s^2 &=&a^2\left[-(1 + 2\Phi)\d \eta^2 + \omega_{i} \d\eta \d x^i + \left((1-2 \Psi)\delta_{i j} + \chi_{ij} \right)\d x^{i}\d x^{j}\right]\,.
\end{eqnarray}
where $\delta_{ij}$ is the spatial metric of the flat background spacetime, with covariant derivative $\bar{\nabla}_i$,  $\Phi$ and $\Psi$ are scalar potentials, $\omega_{i}$ and $\chi_{ij}$ are vector and tensor perturbations.  We have introduced the conformal time, $\eta$, it is related to the coordinate time, $t$, according to $ \d t= a\d \eta $, where $a$ is the scale factor of the universe; its evolution is governed by the Friedmann equation~\cite{2012reco.book.....E}.  Without loss of generality, we will neglect both tensor and vector perturbations for the moment; the impact will be discussed later. 
The four-velocity is given by $[u^0, u^i] = [1 - \Phi ,{ \partial}^{i}v]$, where $v$ is the velocity potential.  
Using equation \eqref{eq:metric},  $\Theta$  has a general form
\begin{eqnarray}\label{eq:therespatial0}
\Theta(\eta, {\x}^i) = \frac{3\HH(\eta)}{a} +\frac{1}{a} \left[ -3\left( \HH(\eta) \Phi(\eta, {\x}^i) + \Phi'(\eta, {\x}^i) \right)+ \nabla^2 v(\eta, {\x}^i)
\right]\,,
\end{eqnarray}
where $v$ is the peculiar velocity potential: $ v^i = \partial^i v$. 
Using  the oi-component of GR
$
\Phi' + \HH \Phi = -4\pi G a^2 \bar{\rho} v = -{2}\Omega_{m} \HH^2 v/{3}\,,
$
and $ \bar{\rho} =  3 H^2 \Omega_m/\kappa$ , it is straight to reduce equation \eqref{eq:therespatial0} to a much simpler form  
\begin{eqnarray}\label{eq:therespatial1}
\Theta(\eta, {\x}^i) = \frac{3\HH(\eta)}{a} +\frac{2}{a}  \Omega_{m} \HH^2(\eta) v(\eta, {\x}^i) +\frac{1}{a}\nabla^2 v(\eta, {\x}^i)\,,
\end{eqnarray}
Irrespective of the gauge choice, $\Theta$ in perturbed FLRW spacetime appears in the form given in equation \eqref{eq:therespatial1}.  It is composed of three terms: the first term is the mean expansion rate, which given by the Hubble rate: $3 \HH/a$, the second term ${a}  \Omega_{m} \HH^2(\eta) v(\eta, {\x}^i) $ as we shall see constrains the size of an under-dense region comoving with the Hubble flow, while the third term  $\nabla^2 v(\eta, {\x}^i)$, which is the divergence of the relative velocity. 
We consider the contribution of the second and third terms separately
\begin{itemize}
\item {\tt{Under-dense region:}} For the under-dense regions, we focus on the first two terms in equation \eqref{eq:therespatial1}.  Using  the Euler equation, that is  $\delta'_{m} = -\partial_{i}\partial^i v$  and expanding the terms in Fourier space gives
\begin{eqnarray}
\Theta(\eta, {\k}) 
&\approx& \frac{3\HH(\eta)}{a}\left[ 1 +\frac{3}{2}\Omega_m \left(\frac{\HH}{k}\right)^2f(\eta)\delta_m({\k},\eta)\right]\,,
\end{eqnarray}
where $f$ is the rate of growth parameter: $f = {\d \ln D_{m}(a)}/{\d \ln a}$, $D_{m}$ is the linear growth factor. On linear scales $\delta_{m}$ is separable: $\delta_{m}(\eta, {\x}) = D_{m}(\eta) \delta_{m}({\x})$.   
 Solving for  $ \left[{\HH}/{k}\right]$ in the limit  $\Theta(\eta, {\k}) \to 0$  gives 
\begin{eqnarray}
 \left[\frac{\HH}{k}\right]= \pm\sqrt{- \frac{2}{3} \frac{1}{\Omega_{m} f(\eta)D_{m}(\eta)} \frac{1}{\delta_{m}({k})}}\,.
\end{eqnarray}
 $ \left[{\HH}/{k}\right]$ is a real only for under-dense regions or voids: $0 <\delta_{m} \le -1 $,  the growth of under-density is bounded, hence, a void the size of the Hubble radius could decouple from the Hubble flow at a finite time in order to maintain causality.

\item {\tt{Over-dense region:}}
For the over-dense region, the important terms are the first and the third terms in equation \eqref{eq:therespatial1}. The Fourier expansion of equation \eqref{eq:therespatial1} in this limit is given by
 \begin{eqnarray}\label{eq:Ellis_Stoeger}
\Theta(\eta, {\k}) \approx \frac{3\HH(\eta)}{a}\left[ 1 -\frac{1}{3}f(\eta)\delta_m({\k},\eta)\right]\,,
\end{eqnarray}
 The expansion vanishes $\Theta = 0$ when $\delta_m(\eta, {\k}_{\star}) = 3/f(\eta)$ is satisfied for over-density within a region $k= 1/R$. In $\Lambda$CDM cosmology, the growth rate is related to the matter density parameter according to $f(z) = \Omega_{m}^{0.54}(z)$.
Ellis and Stoeger presented this argument in \cite{Ellis:2010fr} and interpreted it as the matter horizon indicating the comoving distance beyond which the local matter density cannot catch up with the Hubble flow. 

The argument above is based on cosmological perturbation theory,  it is well-known that the halo model of large-scale structures gives a more physical and beyond perturbation theory description of clustering on small scales~\cite{Cooray:2002dia}.
Therefore, it is consistent to leverage it to estimate the small-scale limit of equation \eqref{eq:therespatial1}.
To achieve this, we made use of the Euler equation  $\delta'_{m} = -\partial_{i}\partial^i v$ to  express $\partial_{i}\partial^i v$ in terms of $ \delta'_{m}$  
$\Theta \simeq 3 H + \partial_{i}\partial^i v
 = 3 H - \delta'_{m}
 \,.
$
With $\Theta$ in this form, we can find the value of the scale factor when and where $\Theta = 0$ in terms of the local matter over-density within the region of space:
$a(\eta_{\star}, {\x}) = a_{\rm{ini}} \exp\left[\left(\delta_{m}(\eta_{\star}, {\x})  - \delta_{\rm{ini}} \right)\right]\,,$
where $a_{\rm{ini}} = a(\eta_{\rm{ini}})$ and $\delta_{\rm{ini}} \ = \delta_{m}(\eta_{\rm{ini}}, {\x}) $.  This is the scale factor beyond which a one-parameter family of time-like geodesics describing gravitationally interacting particles with initial data anchored on an oriented spacetime decouples from the forward-oriented expanding spacetime. 

We would like to estimate $\delta'_{m}$ using the tools of the halo model, so we can estimate the length and time scales when $\Theta=0$. 
For simplicity, we adopt the stationarity approximation in the structure's rest frame. That is, we focus on what an observer at rest in the structure actually sees along null geodesics. Therefore, we express the conformal time derivative of $\delta_{m} $ in terms of the derivative of $\rho$ with respect to the comoving distance
\begin{eqnarray}\label{eq:deltampr}
\delta'_{m} =\frac{\partial \delta_m}{\partial \eta} = \frac{\d r}{\d \eta}\left[-\frac{\partial \delta_m}{\partial r} + \frac{d\delta_m}{d r} \right]\approx - \frac{c}{\bar{\rho}} \frac{ \d \rho}{\d r}  
\approx - \frac{c}{r} \frac{\d \ln \rho}{\d \ln r}\,
\end{eqnarray}
where ${d\delta_m}/{d r} $ is the total derivative along the light cone and ${\partial \delta_m}/{\partial r}$ spatial gradient at fixed time.
We neglect the spatial gradient at a fixed time and consider only the total derivative along the null geodesics since it is what the observer actually sees. 
We have also expressed the matter density contrast in terms of the local matter density $\rho$:
$\delta_{m}  \equiv {\delta \rho}/{\bar{\rho}} = ({\rho - \bar{\rho}})/{\bar{\rho}}\,$ 
leading to 
\begin{eqnarray}\label{eq:halo_Hofz}
\Theta(z,r) \simeq 3 {H}(z) +  \frac{c}{r} \frac{\d \ln \rho}{\d \ln r}(z,r)\,,
\end{eqnarray}
where $c$ is the speed of light.
It is now straightforward  to find the radial distance  when $\Theta=0$
\begin{eqnarray}\label{eq:AH0}
r_{\rm{MH}} = - \frac{ c}{ 3 H} \frac{\d \ln \rho}{\d \ln r}\bigg|_{z=z_{\star}}\,,
\end{eqnarray}
where $r_{\rm{MH}}$ is the matter horizon. 
It is straightforward to estimate ${\d \ln \rho}/{\d \ln r}$ using the NFW halo profile as a function of the radial distance and halo mass at a given redshift~\cite{Diemer:2017bwl}.  The results are shown in Figure \ref{fig:local_gruop} for clusters and in Figure \ref{fig:galaxy_star} for galaxies and stars. 
  \begin{figure}[h]
\centering 
\includegraphics[width=80mm,height=60mm] {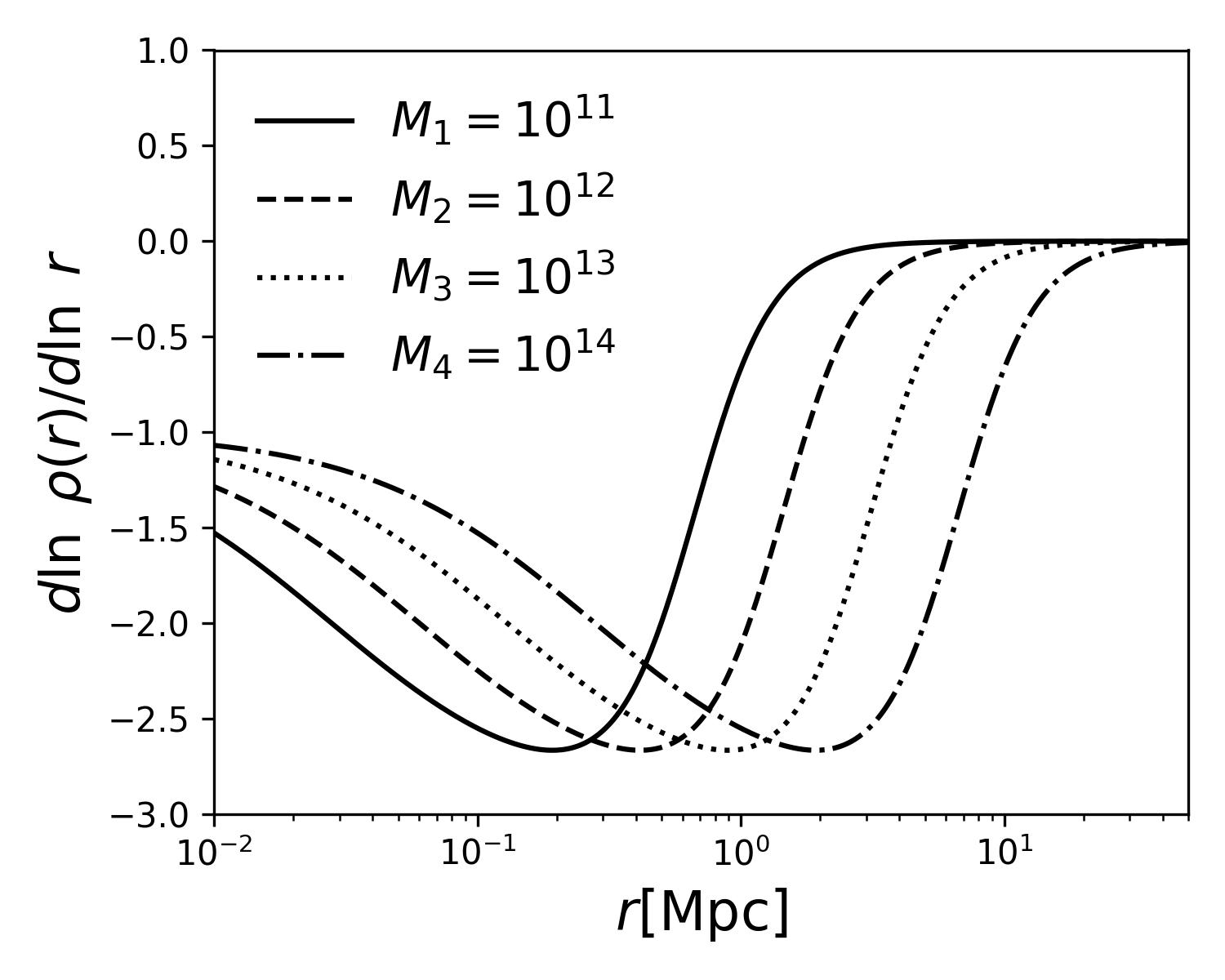 }
\includegraphics[width=80mm,height=60mm] {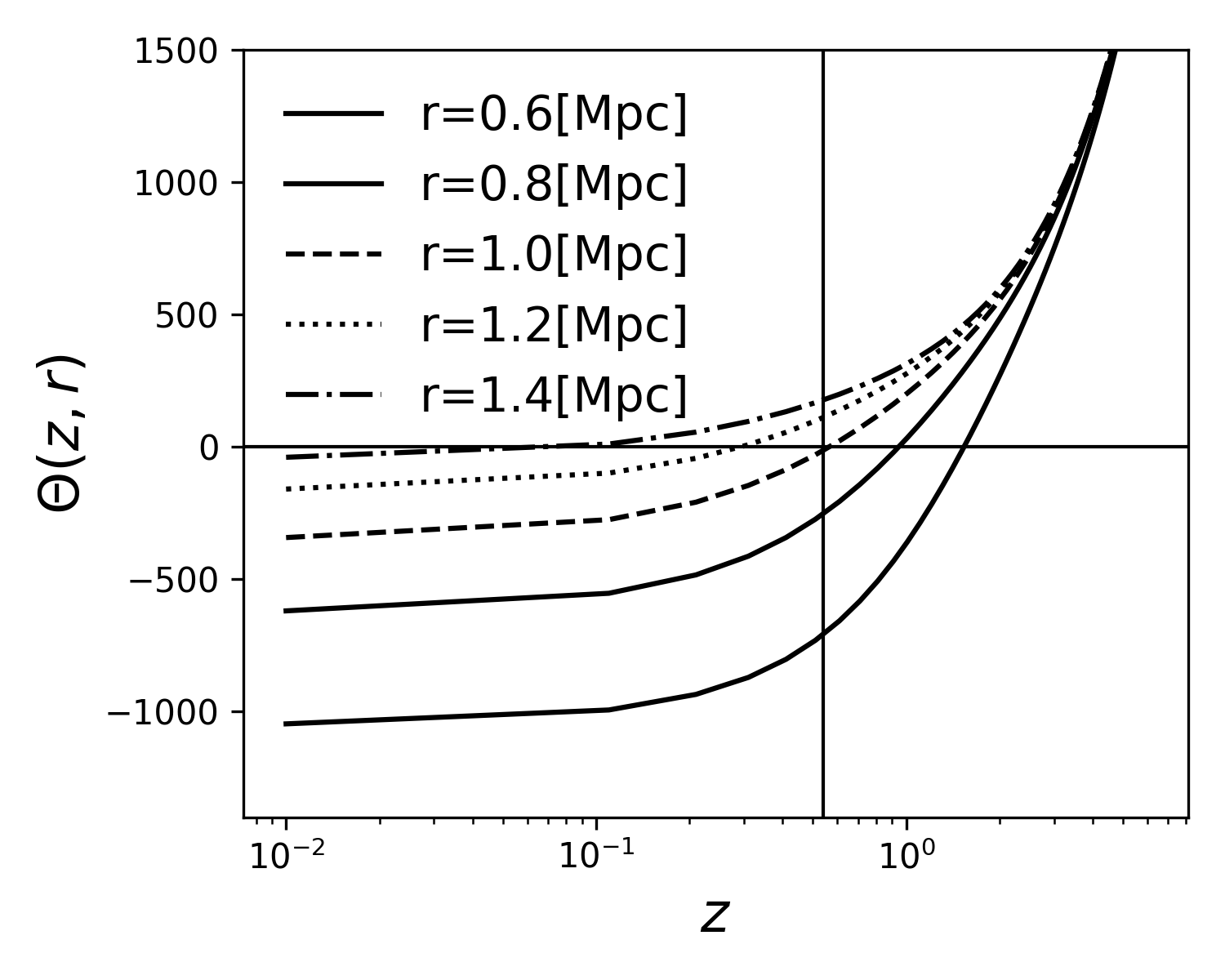 }
\caption{The right panel shows the log gradient of the matter density as a function of radius at $z=0$. The radius corresponding to the minimum log gradient of the matter density is known as the splashback radius of the halo.  It is the first stable orbit a particle will occupy after turnaround~\cite{Diemer:2017bwl}. 
We considered different values of halo mass, with halo concentration set to $c_{\rm{vir}} = 5$.
 The left panel shows the plot of expansion $\Theta$ as a function of redshift at different radial distances with the halo mass set to  $2\times10^{12}M_{\otimes}$. The combination of halo mass and radial distance indicates the critical density at which a clump of matter decouples from the Hubble flow. 
  }
    \label{fig:local_gruop}
\end{figure}
For clusters, we set the halo mass to  $M_{\rm{LG}} = (2\times 10^{12} M_{\otimes}$ corresponding to the mass of a typical cluster such as our Local group~\cite{Li:2007eg,Benisty:2022ive,Sawala:2022ayk}.   The observational constraints on the size of the  Local Group is  $r_{\rm{MH}} \sim(0.95- 1.05) [{\rm{Mpc}}] $ ~\cite{1999AandARv...9..273V, Li:2007eg, Karachentsev:2008st, Kashibadze:2017isc}, 
Given these two observational constraints on mass and radial distance, we can estimate
the corresponding redshift when $\Theta$ vanishes to be $z \sim 0.75$. This indicates the redshift when a sub-region corresponding to our Local group decoupled from the Hubble flow.

Similarly, setting the mass of a typical galaxy such as the Milky Way to $M_{\rm{gal}} = 1.5 \times^{12} M_{\otimes}$, the radius of the Milky Way is about (3-13) kpc~\cite{2012ApJ...759..131B,2024NatAs...8.1302L}, a galaxy of this size and mass decoupled from the Hubble flow at about the redshift of $z \sim 50-90$.
  \begin{figure}[h]
\centering 
\includegraphics[width=80mm,height=60mm] {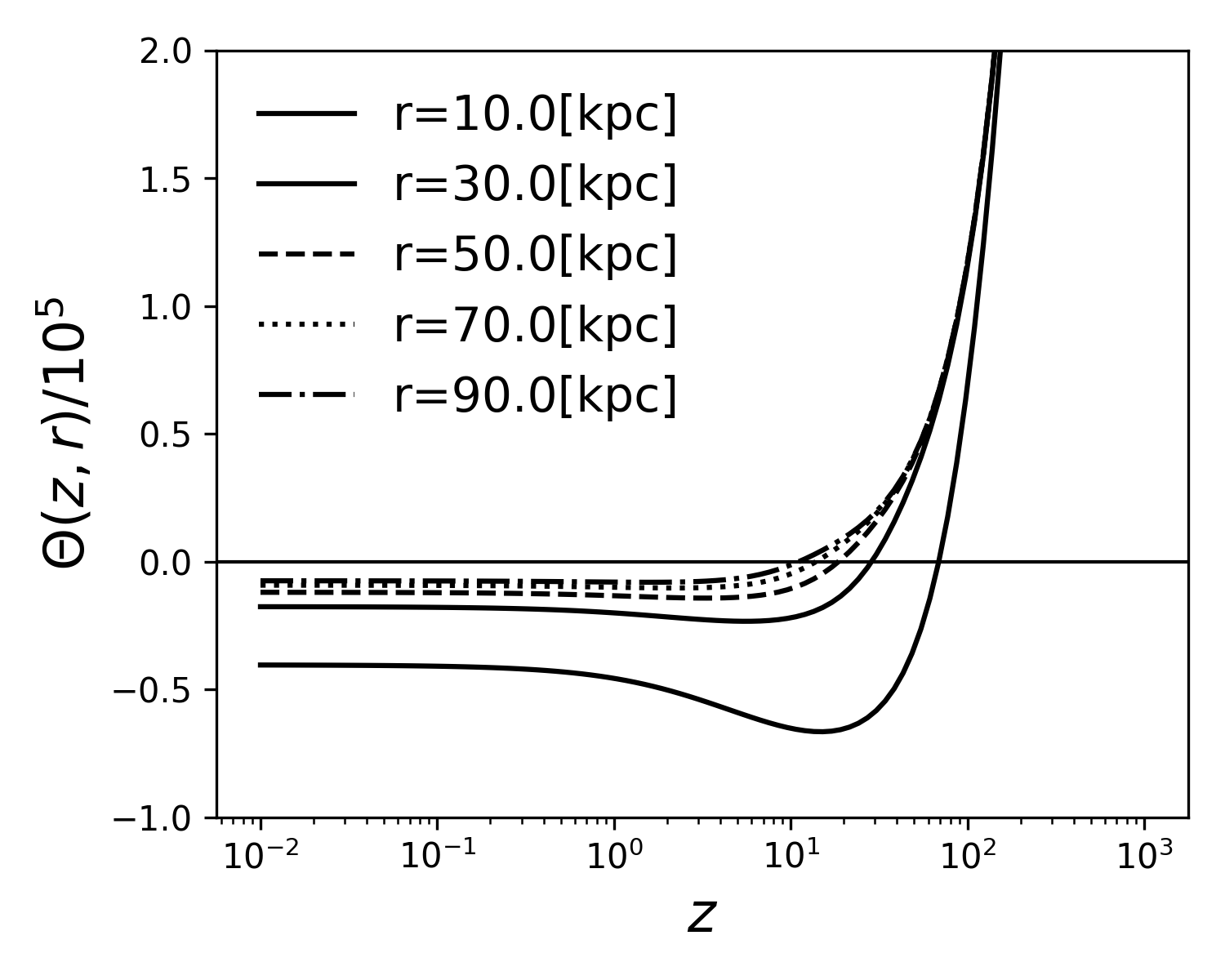 }
\includegraphics[width=80mm,height=60mm] {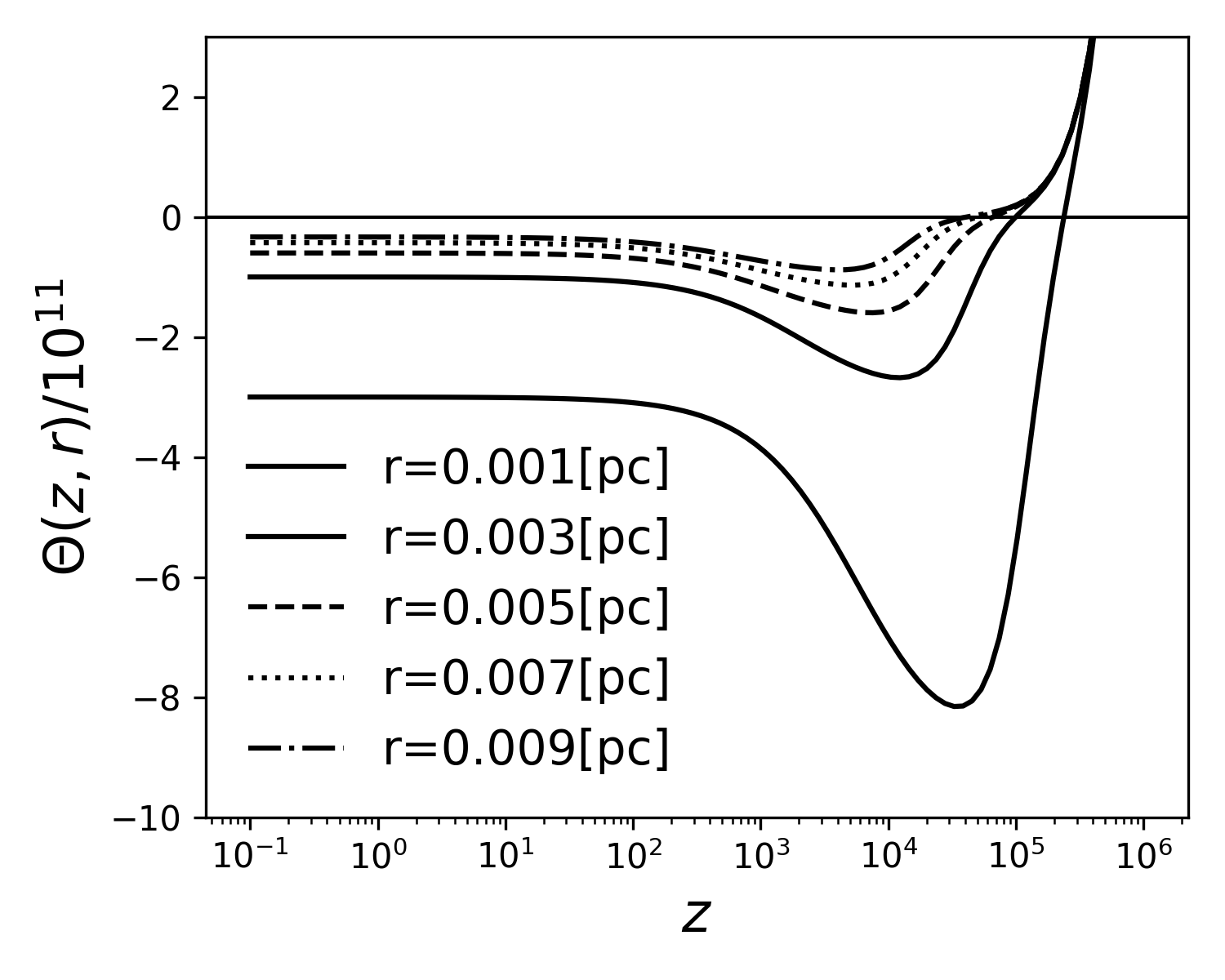 }
\caption{Left panel: This is a plot of $\Theta$ as a function of redshift for a typical galaxy of mass $M_{\rm{gal}} = 1.5 \times^{12} M_{\otimes}$. We considered different radial distances. 
Right panel:  This is a plot of $\Theta$ as a function of redshift for a typical star with mass $M_{\rm{star}} = 2 M_{\otimes} $ .
In both cases,  $\Theta$ changes sign after decoupling and tends to a constant.
}
    \label{fig:galaxy_star}
\end{figure}
 For a typical star, we set the mass to $M_{\rm{star}} = 2 M_{\otimes} $ and consider various radial distances  $(1-9)10^{-3}$pc. A star of this nature decoupled from the Hubble flow at about $z=10^4-10^5$.

\end{itemize}

\subsection{Matter horizon and the global structure of spacetime }\label{sec:breakdown_equivalence}

Horizons in general relativity are defined by the dynamics of non-spacelike integral curves on the spacetime. The non-spacelike integral curves could be timelike or null.  In general, horizons appear in two flavours:
\begin{itemize}
\item Global horizon: these are horizons that emerge because of the constancy of the speed of light. Examples include particle horizons, event horizons.
\item Dynamical horizon: these are horizons that emerge due to the local curvature of spacetime. Examples include the apparent horizon for null integral curves and the matter horizon due to timelike integral curves~\cite{Ellis:2010fr}.  They are dynamical in nature,  that is they evolve as matter and curvature of spacetime change through merger,  accretion and any other viable physical process~\cite{Ashtekar:2003hk}. 
\end{itemize}
Matter horizon in general relativity is an emergent features of spacetime that reflect a breakdown in some of the assumptions or coordinate choice made at the initial  Cauchy surface.  They define boundaries within or limits of predictability under the causal evolution of an initial data.
 To demonstrate this, we consider a small extension of a one-parameter family of geodesics with initial conditions set in the past 
\begin{eqnarray}
\tau  = \tau ' + \Delta\tau \,,
\end{eqnarray}
where  $\tau'  = \tau_{\rm{MH}}$ is the proper time at the matter horizon, i.e $\Theta =0$ hypersurface and $ \Delta\tau$ is a small extension in $\tau$. 
Perturbationg ${\rm{det}}\left[{\mathcal{J}}(\tau)\right]$  slightly around the matter horizon gives
\begin{eqnarray}
{\rm{det}}\left[{\mathcal{J}}(\tau)\right] \approx{\rm{det}}\left[
 {\mathcal{J}}(\tau')\right] \bigg[1 + \frac{1}{ {\rm{det}[{\mathcal{J}}(\tau')]} } \frac{\d {\rm{det}[ \bf{\mathcal{J}}]}}{\d\tau} \bigg|_{\tau =\tau'}\Delta \tau +\frac{1}{2} \frac{1}{ {\rm{det}[{\mathcal{J}}(\tau')]} } \frac{\d^2 {\rm{det}[ \bf{\mathcal{J}}]}}{\d\tau^2} \bigg|_{\tau =\tau'}( \Delta\tau)^2 \bigg]\,.
\end{eqnarray}
 Taking another derivative of equation \eqref{eq:J_theta} and substituting the for  ${\d \Theta}/{\d\tau}$ using the Raychaudhuri equation given in equation~\eqref{eq:expansion_equation}  and implementing the condition for vanishing $\Theta$  gives
    \begin{eqnarray}\label{eq:translation3}
 \frac{1}{ {\rm{det}[{\mathcal{J}}(\tau')]} } \frac{\d^2 {\rm{det}[ \bf{\mathcal{J}}]}}{\d\tau^2} \bigg|_{\tau =\tau'}&=&   -\left[ {\sigma}_{ab}{\sigma}^{ab} 
  + {R}_{ab} {u}^a {u}^b\right] <0\,.
 \end{eqnarray}
Imposing the weak energy condition implies that $ R_{ab}  u^{a} u^{b}\ge0$, and noting that the product of the shear tensors $ \sigma_{ab}\sigma^{ab}>0$ is positive definite implies that everything in the square bracket is positive.  Therefore, for any small extension of geodesics  $\Delta\tau >0$  beyond $\Theta=0$ hypersurface, gives ${\d^2 {\rm{det}[ \bf{\mathcal{J}}]}}/{\d\tau^2}<0$.
That is, any small extension of the geodesic beyond $\tau'$ makes caustics inevitable ~
\begin{eqnarray}
{\rm{det}}\left[{\mathcal{J}}(\tau)\right] \approx{\rm{det}}\left[
 {\mathcal{J}}(\tau_{MH})\right] \big[1-   \frac{1}{2}\left[ {\sigma}_{ab}{\sigma}^{ab} 
  + {R}_{ab} {u}^a {u}^b\right](\Delta \tau)^2\big]
\end{eqnarray}
 in a finite time~\cite{Umeh:2023lbc}.   
  The caustics formation is determined by the local mass density ${R}_{ab} {u}^a {u}^b$ and the strength of the tidal field $ {\sigma}_{ab}{\sigma}^{ab} $. The most important point is that the position of the caustics is in the future of the matter horizon. If a suitable set of coordinates and an affine parameter are made at the matter horizon, the caustics can be avoided. Note that $\Delta \tau >0$ is always positive, see further discussion in the paragraph after equation \eqref{eq:entropy_production}. This is a consequence of the second law of thermodynamics.

\section{Geodesic Flow with Surgery}\label{sec:FLRW_discrete_symmetry}

In this section, we outline the key ingredients for avoiding a gravitational focusing singularity in a manner analogous to the cosmological zoom-in approach. 
The essential idea is to cut the spacetime at the matter horizon and glue it to another sheet of spacetime with opposite orientation. Essentially, we consider two sheets of spacetimes, which are orientation-preserving (${\cal M}^{+})$ and $({\cal M}^{-})$  separated by a boundary $\mathcal{N}$ defined by the matter horizon $\Theta = 0$. 
In this set-up, the coordinates in the neighbourhood of the  point $p_{+} \in M^{+}$ near the boundary $\mathcal{N}$ are related to the coordinates around a point $p_{-} \in M^{-}$,  via   diffeomorphism maps that include orientation-reversing diffeomorphisms (discrete transformations)
\begin{eqnarray}\label{eq:discrte_map}
x^a_{+}   = {-x}^a_{-} +X^a(x) \,.
\end{eqnarray}
In general, the observed universe or the physical reality is described by a union of orientation-preserving manifolds  $\mathcal{M} =  (\mathcal{M}_{+}\setminus \mathcal{D})\cup_{\phi} \mathcal{M}_{-} $, where $\mathcal{D}$ is the excised region.   Again, the boundary $\mathcal{N}$ is a maximal hypersurface.

The equations of GR and Newtonian gravity
are invariant under orientation-reversing discrete transformations of coordinates, such as time reversal (T) and spatial reflection (P). In Newtonian gravitational theory, time reversal $( t  \to  - t )$ changes the direction of velocities $( v^{i} \to  - v^{i})$ but keeps accelerations unchanged: $a^i \to a^i$ (particles fall at the same rate irrespective of time orientation).  
The spatial reflection $x^i \to  - x^i$ keeps the Poisson equation unchanged; hence, T and P independently leave the equations of motion unchanged in Newtonian gravity. 
In general relativity, however,  a combination of time reversal and parity, in the form of PT, is a symmetry of GR. When the geometry is coupled to the external matter fields, the PT symmetry might be enhanced to include charge conjugation, leading to  CPT symmetry (PT symmetry plus charge conjugation, (C) ), which is a fundamental symmetry of the standard model of particle physics~\cite{Luders:1957bpq,Colladay:1996iz}.

However, the impact of the orientation-reversing diffeomorphisms is on the  flow of components of the local coordinates. It is well-known that the local coordinates on the oriented manifolds are chosen via charts. 
In each chart $(\mathcal{U}^{\pm}_{i}, \phi^{\pm}_{i})$, there is a freedom to choose the direction of flow of time and space. 
 For example, for any two instants $ t_1 $ and $ t_2 $, there are two possible directions(forward and backwards) of flow of coordinate time (that is, time in $R^{3,1}$). The forward direction is chosen in standard cosmology, even when the backwards direction is equally likely.  
\begin{eqnarray} \label{eq:time_direction}
   &&\mathcal{M}_{+}\qquad {\rm{ Forward:}} \qquad  t_1 \to t_2 \qquad {\rm{ if}} \qquad \qquad \Delta t  = t_2-t_1>0  \qquad {\rm{for}}\qquad  t :-\infty \to \infty, \\ \nonumber
   && \mathcal{M}_{-}\qquad { \rm{ Backward: }}  \quad t_1 \to t_2\qquad{ \rm{if}}   \qquad \qquad \Delta t =t_2-t_1<0  \qquad   {\rm{for }}\qquad  t: \infty \to -\infty\,. 
\end{eqnarray}
Note that both $\mathcal{M}_{+}$ and $\mathcal{M}_{-}$ are orientation-preserving manifolds, but we have chosen the direction of flow of coordinate time to be opposite to each other. 
 A similar definition exists for the spatial separation; however, special care is required to handle the topology~\cite{Kumar:2023ctp}.  
In general, for an oriented manifold, the atlas $\{(\mathcal{U}^{\pm}_i, \phi^{\pm}_i)\}$ is also oriented, hence for any overlapping charts $(\mathcal{U}^{\pm}_i, \phi^{\pm}_i)$ and $(\mathcal{U}^{\pm}_j, \phi^{\pm}_j)$, the transition function $\phi^{\pm}_j \circ {\phi^{\pm}}_i^{-1}: \phi_i(\mathcal{U}^{\pm}_i \cap \mathcal{U}^{\pm}_j) \to \phi^{\pm}_j(\mathcal{U}^{\pm}_i \cap \mathcal{U}^{\pm}_j)$ has a Jacobian matrix with positive determinant:
$\det\big[ {\partial {x'_{+}}^b}/{\partial x^a_{-}} \big]> 0$ in four dimensions,
where $x^b_{\pm}  = \phi^{\pm} _i(p)$ and $x'^b_{\pm}  = \phi^{\pm} _j(p)$ are coordinates of a point $p^{\pm} \in \mathcal{U}^{\pm}_i \cap \mathcal{U}^{\pm}_j$ in charts $\phi^{\pm}_i$ and $\phi^{\pm}_j$, respectively. The positive determinant constraint ensures that the coordinate transformation preserves the orientation for the entire manifold.

The choice of forward time direction is usually made with an erroneous dismissal of the backwards time direction as being unphysical since we observe an expanding universe~\cite{Lemaitre:1933gd,Nakamura:1998mt}.
This is erroneous because what is measured is the ``proper time" and not the ``coordinate time". 
The specific direction of the flow of coordinate time in equation \eqref{eq:time_direction} is part of the diffeomorphism symmetry of general relativity.
 The proper time, which is the  "actual" time experienced by an object(an observer) between two events along its worldline, is defined through the line element(it depends on the spacetime metric tensor)
\begin{eqnarray}\label{eq:propertime}
\d \tau_{\pm} = \pm \sqrt{\d s^2_{\pm}} = \pm \sqrt{-g^{\pm}_{ab} \frac{\d x^a_{\pm} }{\d \lambda_{\pm}}\frac{\d x^b_{\pm} }{\d \lambda_{\pm}}}\,,
\end{eqnarray}
where $\lambda_{\pm}$ is a parameter, $\d s^2_{\pm}$ is an infinitesimally small distance along the worldline, $g^{\pm}_{ab}$ is is the spacetime metric, $x^a_{\pm}(\lambda_{\pm})$  is a parametrisation of the worldline $\gamma$ and $\pm$ encodes the direction of flow of coordinate time which is due to the Fundamental theorem of algebra. $+$ indicates forward flow of coordinate time, while $-$ indicates backward flow of coordinate time. 
$\tau_{\pm} $ is the time measured by a clock that moves along with an object and not $t$.  Irrespective of the directions of flow of coordinate time, the flow of proper time, $\d \tau_{\pm} $ is asymmetrical in the ambient spacetime. 
Similarly, on any spatial slice, one must make a choice of spatial flow direction: $\Delta x^i = x^i_2-x^i_1$, $\Delta x^i>0$ indicating a choice of forward or positive direction along each coordinate axis and $\Delta x^i<0$ indicating the opposite flow direction. The proper length becomes $ \d\ell  = \pm \sqrt{g_{ij} \d x^i \d x^j}$, the $\pm$ must be chosen so that the proper length remains positive irrespective of the direction of flow. 


The definition of proper time in equation \eqref{eq:propertime} is fully compatible with the  standard FLRW spacetime
\begin{eqnarray}\label{eq:FLRW metric}
\d s^2 = - \d t^2 + a^2(t) \d \Sigma^2\,. \qquad {\rm{where}}\qquad \d \Sigma^2 = \frac{\d r^2}{1-K r^2} + r^2 \left( \d\theta^2 + \sin^2 \theta \d\phi^2\right)\,,
\end{eqnarray}
where $ a~$ is the factor of the universe,  $\d \Sigma^2$ is the spatial part of the metric tensor.  It describes the geometry of the three-dimensional space at a given cosmic time t. It can take three forms depending on the curvature of the universe; $K$ is the curvature constant: $K= [1,0-1]$ indicating positive curvature, flat and open universe.  For the discussion that follows, we focus on $K=0$ universe. Equation \eqref{eq:FLRW metric} is invariant under PT transformation. 
It is also straightforward to show that both directions of the flow of coordinate time (equation \eqref{eq:time_direction}) are consistent with the expanding universe.
The equations of motion for the scale factor of the universe are the Friedmann and Raychaudhuri equations
\begin{eqnarray}\label{eq:Friedmann_equation}
\left(\frac{1}{a} \frac{\d a}{\d t} \right)^2 &=&  \frac{8 \pi G}{3} \rho  
+ \frac{\Lambda}{3} \,,
\\
\frac{1}{a} \frac{\d ^2 a}{\d t^2} &=& - \frac{4 \pi G}{3}\left( \rho + 3 P\right)  + \frac{\Lambda}{3} \,,
\label{eq:Raychaudhuri}
\end{eqnarray}
where $\rho$ is the energy density and $P$ is pressure.  Assuming the equation of state for the perfect fluid,  the pressure is related to the energy density $P =\omega \rho$, where $\omega$ is the equation state of state parameter,  the conservation equation $\dot{\rho} + 3 H \left(\rho + P\right) = 0$ immediately gives $ \rho \sim a(t)^{-3(1+\omega)}$.

 \begin{figure}[h]
\includegraphics[width=80mm,height=60mm]{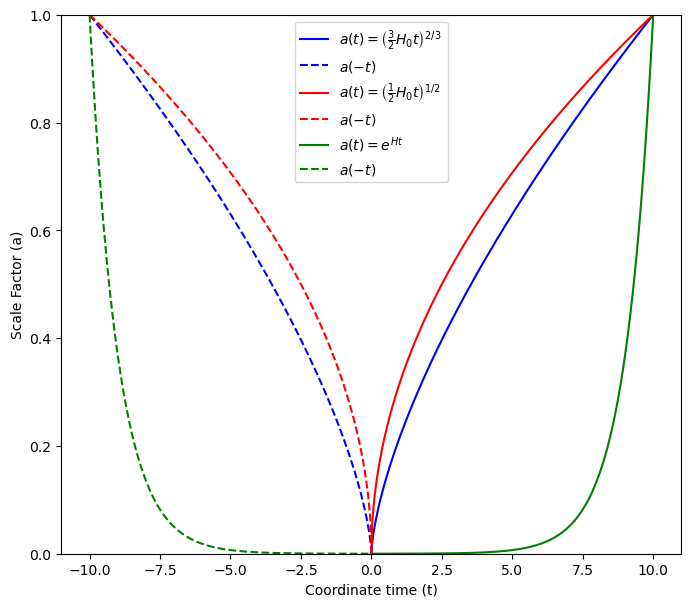}
\caption{
This is a plot of the scale factor ($a(t)$) versus time $(t)$  with $H_0 =1$.  Values of time are chosen to illustrate that the scale factor remains positive irrespective of the direction of flow of coordinate time.  
}
\label{fig:various_epochs}
\end{figure}
According to the standard model of cosmology, the matter content of the universe started off with an Inflaton field slowly rolling down the hill:  $H = \dot{a}/a = H_{\rm{inf}}({\rm{constant}})$  and ended with reheating which launched the radiation dominated epoch with the energy density
 $\rho = \rho_{r0} \left({a_{0}}/{a}\right)^4$ and the matter-dominated epoch with the energy density $\rho = \rho_{m0} \left({a_{0}}/{a}\right)^3$. There is also a late time-dark energy-dominated phase. 
 The inflation era is associated with an exponential rise in the scale factor: $a(t) = a_{\rm{start}} e^{H_{\rm{ini} (t-t_{\rm{start}})}}$, Th inflation ends after about $60$ e-folds$(N =H_{\rm{ini}}(t_{\rm{end}}-t_{\rm{start}})$).
During the radiation era the scale factor grows like $ a = \sqrt{2H_{0}\sqrt{\Omega_{ro}} \left(t-t_1\right) + a_1^2}$ from about $10^{-51}$   to   $7.6 \times 10^{-5}$Gyr, the matter era $
a(t) = \left( \frac{3}{2} H_{0} \sqrt{\Omega_{m0}} \left(t  -t_{2} \right) + a_{2}\right)^{\frac{2}{3}}
$,  from $7.6 \times 10^{-5}$ to 3 Gyr and 
the matter plus dark energy era from 3 Gyr to 13.8 Gyr. The scale factor is normalised to $a_{t_{\rm{today}}} = 1$.

It is clear from figure \ref{fig:various_epochs} that the scale factor(the cosmic clock) increases   $  {\d a(\tau)} > 0$ irrespective of the direction of flow of coordiate time.
Therefore,  irrespective of the choice of the flow of coordinate time(equation \eqref{eq:time_direction}), the direction of flow of proper time is asymmetrical, provided that the appropriate choice of the sign is made in equation \eqref{eq:propertime}.
The thermodynamic arrow of time for the total entropy $ S(\tau) $ of an isolated system is defined in terms of the entropy current \cite{Israel:1976tn,Joshi:2023zve} $ s^a = s u^a +$ (dissipative terms), where the entropy density, $s$ is obtained as its time component. The spatial components correspond to the entropy flux  \cite{Israel:1976tn}.  Focusing on the time component alone, the entropy production rate for an isolated system(the second law of thermodynamics) is given as the divergence of $s^a$:  
\begin{equation}\label{eq:entropy_production}
\nabla_{a} s^a =u^a \nabla_{a} s  + s  \Theta  \ge 0\,.
\end{equation}
where $\Theta = \nabla_{a}u^a $ is the trace of the covariant derivative of a timelike comoving 4-vector, $u^a$ orthogonal to the hypersurface of constant time, and  $s$ is the entropy density. It is related to the total entropy, $S$, according to $s = S/V$,  where $V$ is the volume of the hypersurface.
The covariant derivative along $u^a$, that is $u^a \nabla_{a} = {\d}/{\d\tau}$ relates to the proper time. In general, $\Theta = \d V/ \d \tau /V $  from equation \eqref{eq:J_theta}, hence, the entropy production rate is defined in terms of the proper time $ \d S/ \d \tau \ge 0$. Therefore, the second law of thermodynamics is preserved in both $\mathcal{M}_{+}$ and $\mathcal{M}_{-}$. 

\subsection{Cordinate oreintation and Feynman-Stueckelberg interpretation}\label{sec:Singularity_avoidance}

\begin{figure}[h]
\centering
\includegraphics[width=100mm,height=50mm] {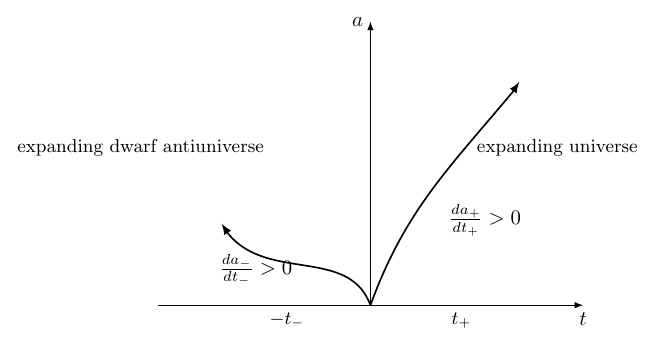}
\caption{
 Two possible time orientations of an FLRW spacetime, both leading to expansion: a universe expanding forward in coordinate time $t_+$, and its coordinate time-reversed conjugate  (anti-universe) is also expanding in reversed coordinate time $t_{-}$.  Though both exhibit $ {da}/{dt} > 0$ in their respective coordinate system,  differences in local spacetime curvature can lead to distinct expansion rates.
} 
\label{fig:dwarf-anti-universe}
\end{figure}

Now that the entropy argument is clarified, we will present details on how to go beyond the matter horizon by imposing discrete symmetry at the boundary $\mathcal{N}$.  
Our approach to avoiding gravitational focusing singularity is to choose the spacetime with an appropriate orientation at the other side of the matter horizon. Specifically, on the background FLRW spacetime $\Theta = 3 H = 3{ \d a}/{\d t }/a$ and the freedom to choose the direction of flow of coordinate time, the  expanding spacetime is described by
\begin{equation}\label{eq:Gexpansion}
\text{Global expansion} \quad \Longrightarrow \quad
\left\{
\begin{array}{ll}
H_{+} =\frac{1}{a_{+}}\frac{ \d a_{+}}{\d t_{+} } > 0, &  \qquad t_{+} : -\infty \rightarrow \infty \\
H_{-} =\frac{1}{a_{-}}\frac{ \d a_{-}}{\d t _{-}} < 0, &  \qquad  t_{-} : \infty \rightarrow -\infty
\end{array}
\right.
\end{equation}
where $a_{\pm} = a(t_{\pm})$. 
 The consequences of the two possible flow directions of coordinate time leading to expanding universes have been discussed earlier in \cite{Kumar:2023ctp,Enrique:2836794,Gaztanaga:2024whs,Gaztanaga:2024vtr} in the quantum gravity context.  It builds on the works of \cite{Schrodinger:1956jnw}.
The standard model of cosmology is built on the choice of the forward flow of coordinate time.  
Similarly, the contraction in both manifolds is defined as
\begin{equation}\label{eq:Gcontraction}
\text{Global contraction} \quad \Longrightarrow \quad
\left\{
\begin{array}{ll}
H_{+} =\frac{1}{a_{+}}\frac{ \d a_{+}}{\d t_{+} } < 0, & \qquad t_{+}: -\infty \rightarrow \infty \\
H _{-}=\frac{1}{a_{-}}\frac{ \d a_{-}}{\d t_{-} } > 0, &  \qquad  t_{-} : \infty \rightarrow -\infty
\end{array}
\right.
\end{equation}
We can now extend this to the local component of the expansion scalar $\Theta$, where $\Theta=\partial_{i} v^i $. 
However, analysing this will require choosing a specific coordinate system to define the orientation of the components of the peculiar velocity, 
For example, in the spherical coordinate system, the component transforms as 
$
(r,\theta,\phi) \to (r,\pi. - \theta,\phi -  \pi)\,, 
$ 
where the radial distance  transofrms as $r \to r$  because it is a positive scalar,  the polar angle $ \theta \to \pi - \theta$  reflects across the origin and   the azimuthal angle $\phi \to\phi -  \pi$ shifts to reflect the inversion. For scalar perturbations, we can use the Euler equation instead to express $\Theta_{\rm{L}}$ in terms of the rate of change of the density contrast
$\delta'_{m} = -\partial_{i}\partial^i v$   and then apply analsysis similar to equations \eqref{eq:expansion} and \eqref{eq:contraction}
\begin{equation}\label{eq:expansion}
\text{Expansion} \quad \Longrightarrow \quad
\left\{
\begin{array}{ll}
\Theta^{+}  =\frac{3}{a_{+}}\frac{ \d a_{+}}{\d t_{+} }-\frac{ \d \delta_{m+}}{\d t_{+} }  > 0, & \qquad   t_{+} : -\infty \rightarrow \infty \\
\Theta^{-} =\frac{3}{a_{-}}\frac{ \d a_{-}}{\d t _{-}}-\frac{ \d \delta_{m-}}{\d t_{-} } < 0, &  \qquad  t_{-} : \infty \rightarrow -\infty
\end{array}
\right.
\end{equation}
and 
\begin{equation}\label{eq:contraction}
\text{Contraction} \quad \Longrightarrow \quad
\left\{
\begin{array}{ll}
\Theta^{+} =\frac{3}{a_{+}}\frac{ \d a_{+}}{\d t_{+} }-\frac{ \d \delta_{m+}}{\d t_{+} } < 0, &  \qquad  t_{+}: -\infty \rightarrow \infty \\
\Theta^{-}= \frac{3}{a_{-}}\frac{ \d a_{-}}{\d t_{-} }-\frac{ \d \delta_{m-}}{\d t_{-} } > 0, & \qquad   t_{-} : \infty \rightarrow -\infty
\end{array}
\right.
\end{equation}
For forward time orientation $ t_{+} : -\infty \rightarrow \infty$, $\Theta  >0$ implies expansion while  $\Theta  <0$ implies collapse to a singularity beyond the matter horizon.  However,  reversing the direction of flow of coordinate time $t_{-} : \infty \rightarrow -\infty$, leads to expansion  $\Theta  <0$, thereby avoiding the singularity.  See figure \ref{fig:dwarf-anti-universe} for further detials.

Finally, we showed in sub-section \ref{eq:particle_formation} that the formation of the matter horizon is associated with the formation of a gravitationally bound system with internal degrees of freedom.
We also showed in sub-section \ref{sec:breakdown_equivalence} that as soon as a gravitationally bound system is formed, it is impossible to extend a one-parameter family of geodesics with initial conditions defined on the same manifold beyond the matter horizon. Any extension of the geodesics is disfavoured by physical considerations such as the equivalence principle, causality,  etc~\cite{Hawking:1991nk,Enrique:2836794,Gaztanaga:2024whs,Gaztanaga:2024vtr}.  
We have also shown in this sub-section that the only physically plausible route to avoiding the focusing singularity is by cutting off the spacetime at the matter horizon and glueing it to another sheet of spacetime with opposite orientation. 
This deduction is consistent with the Feynman-Stueckelberg interpretation in particle physics~\cite{Stueckelberg:1941rg,Feynman:1949hz},  where the unbounded energy catastrophe associated with a negative energy particle moving forward in time is resolved by introducing antiparticles as particles moving backwards in time.  
Here, we resolved the gravitational focusing singularity by describing the internal degrees of freedom of a gravitationally bound particle as a separate universe with coordinate time moving backwards. This allows us to continue the geodesic flow in proper time.

\subsection{Hierarchical structure  formation via piece-wise geodesics}\label{sec:beyond-one}

Sub-section \ref{sec:Singularity_avoidance} provided us with a definition of a gravitationally bound system in an expanding universe as a region of spacetime with coordinate time flowing backwards inside and forward on the outside. Both the inside and outside are separated by the matter horizon. 
  In this section, we use the variational principle to show how this might be included and also illustrate correspondence with the cosmological zoom-in N-body simulation described in sub-section \ref{sec:zoom-in}.
  
\begin{figure}[h]
\centering
\includegraphics[width=70mm,height=50mm] {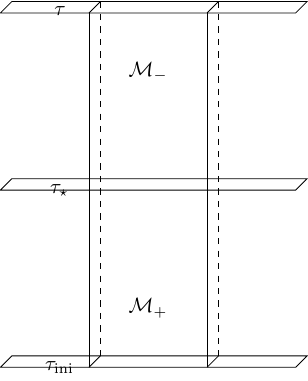}
\caption{A one-parameter family of timelike geodesics with initial condition set at $\tau_{\rm{ini}}$  in $({\cal M}^{+}, g_{ab}^{+})$ with local coordinate time flowing forward, it evolves into the future $\tau_{+} > \tau_{\rm{ini}}$ untill it reaches a maximal hypersurface at  $\tau_{\star}$.  At the maximal hypersurface, it decouples from $({\cal M}^{+}, g_{ab}^{+})$  and turns around (changes orientation) and continues to evolve into the future, $\tau_{-} >\tau_{\star}$ according to an observer at rest in $({\cal M}^{-}, g_{ab}^{-})$ manifold.   }
\label{fig:matching_spacetimes}
\end{figure}

In this set-up, the observed universe is described by a union of orientation-preserving manifolds and the boundary $\mathcal{M} =  (\mathcal{M}_{+}\setminus \mathcal{D})\cup_{\phi} \mathcal{M}_{-} $.
The oriented manifolds on each side of the boundary are endowed with metrics $g_{ab}^\pm$, such that $({\cal M}^{+}, g_{ab}^{+})$ denotes a Lorentian manifold describing an epoch when the initial conditions for a family of nearby geodesics were set on an expanding background spacetime with coooredinate time flowing forward. 
We denote the Lorentian manifold  with the coordinate time orientation reversed as 
$({\cal M}^{-}, g_{ab}^{-})$.
Both manifolds are time-oriented such that the geodesic initialised at $\tau_{\rm{ini}}$ and
evolves on  $({\cal M}^{+}, g_{ab}^{+})$  until it reaches a maximal hypersurface at $\tau_{\star}$ and decouples from the forward flowing coorinate time and continues its subsequent evolution on $({\cal M}^{-}, g_{ab}^{-})$  with the flow of cooredinate time reversed but with the proper time flowing forward. This is illustrated in Figure \ref{fig:matching_spacetimes}.

 The  tangent vectors to the integral curves in $({\cal M}^{\pm}, g_{ab}^{\pm})$ are
\begin{eqnarray}
u^a_{+}
 =\frac{\d x^a_{+}}{\d \tau_{+}}~~~~~~~~~{\rm{and}}~~~~~~~ u_{-}^a     = \frac{\d x^a_{-}}{\d \tau_{-}}   \,,
\end{eqnarray}
where $x_{+}^{a}$ is associated with  the tangent space of ${\cal M}^{+}$ and ${x}^{a}_{-} $ is associated with the tangent space of ${\cal M}^{-}$. 
The energy-momentum tensor obtained from the ensemble of massive particle actions (equation \eqref{eq:massive_particle_actionl}) can  then be written as a sum of massive particle actions on the two sides of the matter horizon
  \begin{eqnarray}\label{eq:piecewise_action}
S (\gamma_{\pm},\gamma'_{\pm}) &=&
 \int_{\tau_{\rm{ini}}}^{\tau_{\star}}
 {L} _{+}\left[ \gamma_{+}( \tau_{+}) , \gamma'_{+}( \tau_{+})\right] \d  \tau_{+} 
 +\int_{\tau_{\star}}^{\tau_{\rm{final}}} 
 {L} _{-}\left[ \gamma_{-}( \tau_{-}) , \gamma'_{-}( \tau_{-})\right] \d  \tau_{-}
 \,,
\end{eqnarray}
where $L_{+}$ is the Lagrangian of the massive particle with initial conditions set on the expanding coordinates with forward flowing coordinate time, $L_{-}$ is the Lagrangian of the massive particle after decoupling from the forward flowing coordinate time,  $\tau_{\star}$ (or the turnaround time or the maximal foliation time, or the time when $\Theta =0$ for a family of massive particles. 
These particles' clocks lose their synchronisation to the forward-flowing coordinate time Hubble flow at a time  $\tau_{\rm{\star}}$. 
 After decoupling from the forward-flowing coordinate time Hubble flow, the massive particles now evolve on the background spacetime with a backwards-flowing coordinate time.

  The critical point of equation \eqref{eq:piecewise_action} with respect to an infinitesimal variation, $s$,  corresponds to an infinitesimal variation of the respective actions
\begin{eqnarray}\nonumber
 \frac{\d S }{\d s} \bigg|_{s = 0}&=&\frac{\d }{\d s} \bigg|_{s = 0} \int_{\tau_{\rm{ini}}}^{\tau_{\star}}
 L _{+}\left[\gamma_{+}( \tau_{+}),{{\gamma}_{+}}( \tau_{+})\right] \d  \tau_{+}
 + 
\frac{\d }{\d s} \bigg|_{s= 0} \int_{ \tau_{\star}}^{\tau_s}
 L _{-}\left[\gamma_{-}( \tau_{-}),{{\gamma}_{-}}( \tau_{-})\right] \d  \tau_{-}\,.
\label{eq:vary_Null_action}
\end{eqnarray}
Performing the functional derivative of the Lagrangian and  imposing proper variation at the final  endpoints ($\xi^i(\tau_{\rm{ini}}) = \xi^i(\tau_{\rm{final}}) =0$) of the geodesic gives 
\begin{eqnarray}
0 &=& \left[ 
 \frac{\partial L _{+}}{\partial {{\gamma}_{+}^i} }(\tau_{\star})\xi_{+}^i( \tau_{\star}) - \frac{\partial L _{-}}{\partial {{\gamma}^i_{-}} }(\tau_{\star})\xi_{-}^i( \tau_{\star})  \right]
\\ \nonumber &&
+\int^{\tau_\star}_{\tau_{\rm{ini}}}  \left(\frac{\partial L_{+}}{\partial \gamma_{+}^i}( \tau_{+} )
- \frac{\d}{\d  \tau_{+}} \frac{\partial L_{+}}{\partial {{\gamma}_{+}^i} }(\tau_{+})\right)\xi_{+}^i( \tau_{+}) \d  \tau_{+} 
+ \int^{\tau_{\rm{final}}}_{\tau_{\star}}  \left(\frac{\partial L_{-}}{\partial \gamma_{-}^i}( \tau_{-}) - \frac{\d}{\d  \tau_{-}} \frac{\partial L_{-}}{\partial {{\gamma}_{-}^i} } (\tau_{-}) \right)\xi_{-}^i(\tau_{-})  \d  \tau_{-}\,,
\end{eqnarray}
where $\xi^a_{\pm} $ is a deviation vector $\xi^a_{\pm} = {\partial x^a_{\pm}(\tau,s)}/{\partial s}\,. $
We require that the curves are piece-wise smooth at the boundary, $\tau_{\star}$: Using equation \eqref{eq:discrte_map}, it becomes clear that  $\xi^i_{-}(\tau_{\star}) = -\xi_{+}^i (\tau_{\star})$. 
Mathematically,  this translates to the requirement that the Euler-Lagrange equations are independently  satisfied~\cite{Markvorsen:2022arXiv220713515M,Umeh:2023lbc}
\begin{eqnarray}\label{eq:EL_Null_A}
  \frac{\d}{\d  \tau_{+} } \frac{\partial L_{+}}{\partial {{\gamma'}_{+}^i} }-\frac{\partial L_{+}}{\partial \gamma_{+}^i} &=&  0 
\qquad {\rm{for}}\quad   \tau_{+} \in [\tau_{\rm{ini}}, \tau_{\star}] \,,
 \\
   \frac{\d}{\d  \tau_{-}} \frac{\partial L_{-}}{\partial {{\gamma'}_{-}^i} } -\frac{\partial L_{-}}{\partial \gamma_{-}^i}&=&  0 
  \qquad {\rm{for}} \quad  \tau_{-} \in [\tau_{\star},\tau_{\rm{final}}] \,.
  \label{eq:EL_Null_C}
\end{eqnarray}
And at the boundary we have 
\begin{eqnarray}\label{eq:Israel}
 \left[     \frac{\partial L_{+}}{\partial {{\gamma}_{+}^i} }(\tau_{\star}) +  \frac{\partial L_{-}}{\partial {{\gamma}_{-}^i} }(\tau_{\star}) \right]\xi^i_{+}( \tau_{\star})  = 0\,.
\end{eqnarray}
Equation \eqref{eq:Israel} is a generalised junction condition. It reduces to  Israel junction condition in appropriate limits~\cite {Israel:1966rt}. 
\begin{itemize}
\item {\tt{Geodesics}}: We consider the limit of a massive particle Lagrangian
\begin{eqnarray}\label{eq:point_particle}
L_{\pm}(x^a_{\pm},{x'}^a_{\pm}) =-m_{\pm}  \sqrt{-g^{{\pm}}_{ab} \frac{\d x^a_{\pm} }{\d\tau_{\pm}}\frac{\d x^b_{\pm}}{\d\tau_{\pm}} } 
 =  -m_{\pm} \sqrt{-g^{{\pm}}_{ab}  u^a_{\pm}u^{b}_{\pm} }\,.
\end{eqnarray}
Putting equation \eqref{eq:point_particle} in equations \eqref{eq:EL_Null_A} and \eqref{eq:EL_Null_C} gives the respective geodesic equations
\begin{eqnarray}\label{eq:geodesic_equation}
u_{+}^a \nabla_a u^{b}_{+} = 0\,,\qquad  {\rm{and}}\qquad  u_{-}^a \nabla_a u^{b}_{-} = 0\, .
\end{eqnarray}
 The boundary conditions for the geodesics are obtained by putting the Lagrangian in the generalised Israel junction condition (equation \eqref{eq:Israel})
\begin{eqnarray}\label{eq:four_velocity}
 \left[  \frac{1}{L_{+}}  \frac{\d x_{+}^a }{\d \tau_{+}}(\tau_{\star}) +\frac{1}{L_{-}}\frac{\d x_{-}^a }{\d \tau_{-}}(\tau_{\star})  \right]\xi^i_{+}(\tau_{\star}) \approx \left[  \frac{1}{L_{+}} u^a_{+}\bigg|_{\mathcal{N}}+\frac{1}{L_{-}}u^a_{-}\bigg|_{\mathcal{N}}\right]\xi^i(\tau_{\star}) = 0 \,.
\end{eqnarray}
For massive particles ${L_{\pm}}=- m_{\pm}$, hence $u^a_{+} \big|_{\mathcal{N}}+u^a_{-}\big|_{\mathcal{N}} = 0$.

\item {\tt{Geodesic deviation equation}:}

The Lagrangian for the gravitational field felt by a family of massive particles is given by
\begin{eqnarray}\label{eq:Lagrangian_Null_GDE}
L\left[\xi^a_{\pm},\frac{\d {\xi}^a_{\pm}}{\d\tau_{\pm}}\right] =  \frac{1}{2}\eta_{ab} \frac{\d \xi^a_{\pm}}{\d \tau_{\pm}}  \frac{\d \xi^b_{\pm}}{\d \tau_{\pm}}  - \frac{1}{2} R^{\pm}_{abcd} u^a_{\pm} u^b_{\pm} \xi^b_{\pm} \xi^d_{\pm} \,.
\end{eqnarray}
Putting equation \eqref{eq:Lagrangian_Null_GDE}  in equations \eqref{eq:EL_Null_A} and \eqref{eq:EL_Null_C} gives the geodesics deviation equations for both $\mathcal{M}_{-}$ and $\mathcal{M}_{+}$
  \begin{eqnarray}\label{eq:Jacobieqn1}
     \frac{\d^2 \xi^a_{+}}{\d {\tau}^2_{+}} = - R^a_{+}{}_{cbd} u^c_{+} u^d_{+} \xi^b_{+}\,,
~~~~~~~~~~~~~~~~~~~~
      \frac{\d^2 \xi^a_{-}}{\d {\tau} ^2_{-}} &=& - R^a_{-}{}_{cbd} u^c_{-} u^d _{-}\xi^b_{-} \,,
  \end{eqnarray}
Furthermore, putting  equation \eqref{eq:Lagrangian_Null_GDE} in the generalised Junction condition gives the junction condition for the second fundamental form
\begin{eqnarray}\label{eq:Second_fundamental_form}
\left[  \frac{\d \xi^a_{+}}{\d   \tau_{+}}( \tau_{\star})-\frac{\d \xi^a_{-}}{\d   \tau_{-}}( \tau_{\star}) \right] \xi_{a} = 0
 = \left[  \nabla_{a} u^{+}_{b}\bigg|_{\mathcal{N}}-\nabla_{a} u^{-}_{b}\bigg|_{\mathcal{N}}\right]\xi^{a} \xi^{b} 
 = \left[  K_{ab}^{+}-K_{ab}^{-}\right]\xi^{a} \xi^{b}\,,
\end{eqnarray}
  where $K_{ab}^{\pm}$ is the extrinsic curvature tensor on the hypersurface.
   We made use of $  {{\d} \xi^{a}_{\pm}}/{{\d} \tau}   =  \nabla^{b} u^a_{\pm} \xi^b_{\pm} $, The deviation vector on each side is  orthogonal to respective  $u^a_{\pm}$: $u^{\pm}_{b}\xi^b_{\pm}  =0$.
Using the  decomposition of respective extrinsic curvature tensors$K^{\pm}_{ab} $ is given by  $K^{\pm}_{ab} = \frac{1}{3}\Theta^{\pm} h^{\pm}_{ab }+\sigma^{\pm}_{ab}$.
  We can obtain the respective propagation equations on each side of the boundary
\begin{eqnarray}
\frac{ {\rm{D}} {{\Theta_{\pm}}} }{{\rm{D}} \tau_{\pm}} &=& - \frac{1}{3}{\Theta_{\pm}^2} - {\sigma}^{\pm}_{ab}{\sigma}^{ab}_{\pm}
- {R}^{\pm}_{ab} {u}^a_{\pm} {u}^b_{\pm}\,,
\label{eq:expansion_equationpm}
\\
\frac{ {\rm{D} } {{\sigma}}^{\pm}_{ab}}{{\rm{D}} \tau_{\pm}} &=& - \frac{2}{3} {\Theta_{\pm}} {\sigma}^{\pm}_{ab} - {\sigma_{\pm}}^{c}{}_{\<a}{\sigma_{\pm}}_{b\>c} 
 - {C}^{\pm}_{acbd}{u}^c_{\pm} {u}^d_{\pm}\,.
\label{eq:shear_equationpm}
\end{eqnarray}

At the boundary  layer, the key requirement is that expansion vanishes and the rate of shear deformation tensor is continuous at the boundary  
  \begin{equation}\label{eq:vanishing_expansion}
 \Theta^{\pm}\bigg|_{\mathcal{N}}  = 0\,.  \qquad \qquad \sigma^{+}_{ab} =- \sigma^{-}_{ab} 
\end{equation}   
According to \cite{Umeh:2023lbc}, the vorticity tensor could be generated by the dynamics at the boundary layer. 
 The initial conditions for $\Theta$ in the $-$ phase are set $\Theta_{-\rm{ini}}=0$.
\end{itemize}

Finally, the lagrangian for the energy-momentum tensor for the ensemble of massive particles is constructed from the piece-wise action for massive particles given in equation \eqref{eq:piecewise_action}
\begin{eqnarray}\label{eq:matter_action}
{L}_{\rm{M}}& =& \sum_{\ell}S_{\ell}\left(\gamma^a_{+}(\tau_{\ell}),{\gamma^a_{+}}'(\tau_{+\ell}) \right) \frac{\delta^{(4)}\left(x^a_{+} - \gamma^a_{+\ell}(\tau_{+\ell})\right)}{\sqrt{-g_{+}(x^{a}_{+})}}
 +\sum_{\ell}S_{\ell}\left(\gamma^a_{-}(\tau_{-\ell}),{\gamma^a_{-}}'(\tau_{-\ell}) \right) \frac{\delta^{(4)}\left(x^a_{-} - \gamma^a_{-\ell}(\tau_{-\ell})\right)}{\sqrt{-g_{-}(x^{a}_{-})}}\,,
\end{eqnarray}
where the Dirac delta function $\delta^{(4)}\left(x^a_{\pm} - \gamma^a_{\pm\ell}(\tau_{\pm\ell})\right)$ ensures that the action is evaluated at the location of the particle.  The energy-momentum tensor is given by
\begin{eqnarray}\label{eq:MatterEMT}
T^{+}_{ab} &=& \sum_{\ell} { m_{\ell}}\int_{\tau_{\rm{ini}}}^{\tau_{\star}}\d\tau_{+} \frac{\d \gamma^a_{+} }{\d\tau_{+}}
 \frac{\d \gamma^b_{+}}{\d\tau_{+}} 
\frac{\delta^{4}\left(x^a_{+} - \gamma^a_{+\ell}(\tau)\right) }{\sqrt{-g_{{+}}(x^{a}_{+})}}\,,
\ \\
T^{-}_{ab} &=& \sum_{\ell} { m_{\ell}}\int_{\tau_{\star}}^{\tau_{\rm{final}}} \d\tau_{-}\frac{\d \gamma^a_{-} }{\d\tau_{-}}
 \frac{\d \gamma^b_{-}}{\d\tau_{-}} 
\frac{\delta^{4}\left(x^a_{-} - \gamma^a_{-\ell}(\tau_{-})\right) }{\sqrt{-g_{{-}}(x^a_{-})}}\,.
\end{eqnarray}

In comparison to the Cosmological zoom-in N-body simulation, the low resolution simulations correspond to particle propagation in $({\cal M}^{+}, g_{ab}^{+})$, while the high resolution region of inteest corresponds to phase of particle propagation in $({\cal M}^{-}, g_{ab}^{-})$. The more explicit example on the build-up of hierarchy is given in sub-section \eqref{sec:discusion}.  The region of interest is defined dynamically by the matter horizon. The boundary conditions for the tidal field are given in equation \eqref{eq:vanishing_expansion}. In a cosmological zoom-in N-body simulation,  it is customary to go back in time to sample the initial conditions of the particles that will form the zoom-in region. Equation \eqref{eq:focusing_theorem2} shows that such a region can easily be identified given the local expansion scalar.  

\subsection{Glueing of both spacetimes and Virial theorem }\label{eq:spacelike_bridge}

When two manifolds are glued at a boundary in GR, the metric on the hypersurface must be continuous at the boundary. Without loss of generality, we focus on the spacelike hypersurface for bravity.
Form equation \eqref{eq:metric}, the metric on the hypersurface of constant proper time $\Sigma: \tau_{+}(t,x^i) =$ const. is given by
\begin{eqnarray}\label{eq:spatialmatric}
\d s^2_{+}\big|_{\Sigma_{+}} \approx a^2_{+}(\tau) \left[ \big(1-2 \Psi_{+}\big)\delta_{ij}  \right]{\d }x^{i}_{+}
{\d} x^{j}_{+}\,.
\end{eqnarray}
where the corresponding expression in $\mathcal{M}_{-}$ universe can essily be derived in the same gauge. 
 Sometimes, it is mathematically more convenient to work in a translated proper time coordinate system,  $\tilde{\tau}=  \tau -\tau_{\rm{MH}}$, where the boundary hypersurface, $\mathcal{N}$ is then located at $\tilde{\tau}=0$ so that    $ \tilde{\tau}  >0$  parameterises geodesics in  $\mathcal{M}_{-}$ while  $\tilde{\tau}<0$ parameterises the geodesics  $\mathcal{M}_{+}$.
It is immediately clear that the metrics on the hypersurfaces $h^{\pm}_{ab} =g^{\pm}_{ab} + u^{\pm}_a u^{\pm}_b$  are conformally Euclidean
$h^{+}_{ab} = \Omega_{+}^2 \delta_{ab}^{+}$,  and  $ h^{-}_{ab} = \Omega_{-}^2 \delta_{ab}^{-}
$, 
where $\Omega^2_{\pm} =a^2_{\pm}  \big(1-2 \Psi_{\pm}\big)\ $ are the effective conformal factors.   Note that the diffeomorphism transformation changes the metric structure, so the evolution $\Phi_{-}$ is different from that of $\Phi_{+}$
The metrics have the form of the standard FLRW spacetime (see equation \eqref{eq:FLRW metric}), hence they describe separate universes, 
This is consistent with the separate universe argument in the context of structure formation in the universe, where overdense regions of the universe are treated as separate, locally homogeneous and isotropic mini-universes with different effective cosmological parameters~\cite{Dai:2015jaa,Desjacques:2016bnm,Umeh:2019qyd,Umeh:2019jqg}.
 Since both  $h^{-}_{ab} $  and $h^{+}_{ab} $, are conformal to the  Euclidean  metric $\delta_{ij}$ which has zero expansion, we can set  $ \delta_{ab}^{-} = \delta_{ab}^{+}$, which then implies that 
\begin{eqnarray}\label{eq:conformalmap_space}
h^{-}_{ab}  =  \frac{ \Omega_{-}^2}{ \Omega_{+}^2}h^{+}_{ab}  =  \Omega^2_{S} h^{+}_{ab} \,,
\end{eqnarray}
where $\Omega_{S} = { \Omega_{-}}/{ \Omega_{+}}$ is the ratio of the effective scale factor in the universe moving backwards in time to the scale factor of the universe moving forward in time.  Moreover, note that this continuity condition is different from the Darmois-Israel condition, which requires that the metric on the hypersurface satisfy a trivial continuity condition $h_{ab}^{+} \stackrel{\Sigma}{=} h_{ab}^{-} $ \cite{Darmois1927, Israel:1966rt}.  Equation \eqref{eq:conformalmap_space} reduces to the Darmois-Israel condition in the limit $ \Omega_{S} =1$. The consequence of the form of the hypersurface metric condition (equation \eqref{eq:conformalmap_space}) on the curvature tensor is presented elsewhere~\cite{Umeh:2025pm}. The essential difference is that the decomposition of the Riemann tensor in the distributional sense with a step function becomes ill-defined; however, the variational calculus approach remains consistent, as the discussion in sub-section \ref{sec:beyond-one} shows. 
Again, more technical details of this will be presented in ~\cite{Umeh:2025pm}, where it is shown that the hypersurface  $\Sigma$  is no longer a single coherent surface layer, but rather two different geometric surfaces that are identified point-by-point as we show below.

The necessary condition for glueing both metrics at the matter horizon is that points on both spacelike hypersurfaces $\Sigma_{-,\tau}$  and   $\Sigma_{+ ,\tau}$ are  connected  by a smooth map
$
\varphi: \Sigma_{-,\tau} \to \Sigma_{+,\tau} \,,
$
where $\varphi$ is a diffeomorphism that maps points $P_{-} \in  \Sigma_{-,\tau}$, to points $P_{+} \in  \Sigma_{+,\tau}$ (recall equation \eqref{eq:discrte_map}, 
${x}^{a}_{-} \to - x_{+}^{a} + X^{a}$)
where  $x_{-}^{a} $ denotes the local coordinates in  the tangenet space of $\mathcal{M}_{-}$  and ${x}^{a}_{+} $  is the local coordinates in   the tangenet space of $\mathcal{M}_{+}$. 
Under this transformation, the metric on the spacelike hypersurface  transforms as 
\begin{eqnarray}\label{eq:metrictransform}
{h}^{+}_{ab}  - {h}^{-}_{ab}  = \left[X^{c} \partial_{\bot c} {h}^{+}_{ab} + {h}^{+}_{ab} \partial_{a} X^{c} + {h}^{+}_{ac} \partial_{\bot b} X^{c} \right]\,,
\end{eqnarray}
where  $X^a$ is a vector field whose exact form will be specified shortly. 
The terms in the square brackets are proportional to the Lie derivative of a spacetime metric tensor  
$
\mathcal{L}_{X} h^{+}_{ab} = X^c \partial_{c} h^{+}_{ab} + h^{+}_{cb} \partial_{a}X^c + h^{+}_{ac}\partial_{b} X^c
  = \nabla_{a} X_{b} + \nabla_{b}X_a\,.
$
Using equation \eqref{eq:conformalmap_space} in equation \eqref{eq:metrictransform} lead to a  conformal Killing equation 
 \begin{eqnarray}\label{eq:CKE} 
 \mathcal{L}_{X} {{h}^{+}_{ab}}(x^{c}) 
= 2 \phi(x^{a}){{{h^{+}_{ab}}}}(x^{c}) \,,
\end{eqnarray}  
 The $X^{a}$  that solves equation \eqref{eq:CKE} is  called the conformal Killing vector field of $h^{+}_{ab} $ and $\phi(x^{c}) =( 1-\Omega^2_{S}(x^{c}) )/2$ is a redefined conformal factor.  
 Because the metric  $h^{+}_{ab} $   is conformal to the Euclidean metric : 
 ${h}^{+}_{ab} =\Omega_{+}^2 \delta^{+}_{ab}$, we can further express equation \eqref{eq:CKE}  in terms of the Euclidean metric. 
Acting on this with the Lie derivative gives $ \mathcal{L}_{X} {{h}^{+}_{ab}} =\Omega^2_{+}(\eta) \mathcal{L}_{X} {{\delta}^{+}_{ab}} + 2\Omega_{+}(\eta)X^{x}\nabla_{x}\Omega_{+}(\eta){{\delta^{+}_{ab}}}$. Putting this in equation \eqref{eq:CKE}, transforms to another conformal Killing equation with a different conformal factor
 \begin{eqnarray}\label{eq:CKE2nd}
 \mathcal{L}_{X} {{\delta}^{+}_{ab}}(x^{c}) 
= 2 \phi'(x^{c}){\delta_{+{ab}}} (x^{c}) \,,
\end{eqnarray}  
where $ \phi'(x^{c})   \equiv  \left[ -X^{b}\nabla_{b}\Omega_{+}(\eta)/\Omega_{+}(\eta) +   \phi(x^{c}) \right]$. 
We fix the value of $ \phi'(x^{c})$ by taken the trace and substituting  back in equation \eqref{eq:CKE2nd} 
\begin{eqnarray}\label{eq:CKE3}
  \partial_{a}X_{b}+\partial _{b}X_{a}=\frac{2}{3}{\delta}_{ab}\partial_{c}X^{c} \,,
\end{eqnarray}
where the redefined conformal factor is given by $\phi'(x^{c})  = \partial_{c}X^{c} /3$, 
The $ {X}^{b}$  that solves equation \eqref{eq:CKE3} is given by \cite{levine1936}
 \begin{equation}\label{eq:symmetryoffLSS}
 {X}^{b} = \alpha^{b} + M^{b}{}_{a} x^{a}_{+} + \lambda x^{b}_{+} + 2 \left(x_{{+}a} \beta^{a}\right) x^{b}_{+} - x_{{+}a} x^{a}_{+} \beta^{b} \,,
 \end{equation}
 where $\alpha^{a}$, $\beta^a$ and $\lambda$ are  spatially constant.  $M^{a}{}_{b} $ and $\alpha^{a} $ are associated with rotations and translations respectively.  $\lambda $ is associated with dilatation and $\beta^{b}$ is associated with the special conformal transformation. 
Equation \eqref{eq:symmetryoffLSS} is anti-conformal transformation
$
 {\rm{det}} \left[{\partial x_{+}^2}/{\partial x_{-}^2}\right] = \left(-1 +  \lambda + 2 x^a_{+} \beta_a\right)^3\,.
$
 The effective conformal factor can be expressed in terms of $X^{i}$ as
\begin{eqnarray}\label{eq:scale_factor}
\Omega^2_{S} = 1 - 2 \frac{\mathcal{L}_{X} \Omega_{+}}{\Omega_{+}} - 2 \phi'(x) = 1 - 2 \frac{\mathcal{L}_{X} \Omega_{+}}{\Omega_{+}} - \frac{2}{3} \left(\lambda + 2 x^{i}_{+} \beta_{i}\right)\,,
\end{eqnarray}
where $\phi'(x^a_{+})= \partial^{a} X_{a} /3=  \left(\lambda + 2 x^{i}_{+} \beta_{i}\right)/3\,.$   

The fact that the diffeomorphism transformation(equation \eqref{eq:discrte_map},) reduces to a global symmetry for scalar perturbations in the neighbourhood of the boundary has important physical consequences. We will focus on the Dilatation symmetry, which is responsible for the stability of the gravitating system of particles. In Newtonian gravity,  the Virial theorem is imposed to prevent a collapse to a singularity after turnaround. 
Here, we derive the Virial theorem directly from the Dilatation symmetry.
Noether's theorem states that any global symmetry is associated with a conserved quantity.  The Dilatation symmetry has a conserved quantity associated with it (Dilatation current or virial function )
$G = - \sum_{i} {\bf{x}}_{i} \cdot {\bf{p}}_{i}\,,$
where ${\bf{p}}_{i}$ is the momentum of the i-th particle.  
Therefore, the time derivative of $G$ gives 
\begin{eqnarray}\label{eq:Dcurrent}
\frac{\d G}{\d t}  =  2 T +  \sum_{i} {\bf{x}}_{i} \cdot {\bf{F}}_{i} \,,
\end{eqnarray}
where ${\bf{F}}_{i} =  {\d} {\bf{p}}_{i}/\d t $  and $T = \sum_{i} p^2_{i}/m_{i} /2$ is the kinetic energy. The  force fields that arise from a potential $V({\bf{x}}_{i} )$ transforms as $V(\lambda{\bf{x}}_{i} )  = \lambda^nV({\bf{x}}_{i} )$ and for  $ {\bf{F}}_{i}  = - \nabla_{i} V$,  equation \eqref{eq:Dcurrent} becomes   ${\d G}/{\d t}  = 2 T - n V$, where   ${\bf{x}}_{i} \cdot {\bf{F}}_{i} = - nV$.  Since $G$ is a constant of motion for exact dilatation symmetry ${\d G}/{\d t} =0$: $2T  = n V$.
The dilatation symmetry we derived in equation \eqref{eq:symmetryoffLSS} is not exact because we neglected tensor and vector perturbations; therefore, for a stable bound system,  the Virial theorem holds on average,  that is, the time average of equation \eqref{eq:Dcurrent} over a sufficiently long period $T$ tends to zero:
\begin{eqnarray}
\bigg\<\frac{\d G}{\d t} \bigg\> = \lim_{\tau \to 0}\frac{1}{\tau } \int_{0}^{\tau} \frac{\d G}{\d t}  \d t = 0\,.
\end{eqnarray}
There, we recover the Virial theorem  $\<2T\> = n \<V\>$. In the inverse square law limit $n=1$(Newtonian gravity).

\subsection{Hierarchical  Cosmological zoom-in perturbation theory}\label{sec:discusion}

The matter in the universe is organised in a hierarchical ``bottom-up" fashion, smaller structures formed first and then merged and grew to create larger ones. This process is the cornerstone of the $\Lambda$CDM  model of cosmology.. Using the details provided in sub-section \ref{sec:beyond-one}, we describe how the internal dynamics of these structures, such as galaxies, clusters, etc, can be described within the expanding universe. 
\begin{itemize}
\item {\tt{Light elements scale:}} The light elements, such as hydrogen, helium, etc., formed at the epoch of nucleosynthesis when point particle approximation is valid down to the scale of the fundamental particles (the density fluctuations were of the order of 1 part in 100,000) . These particles merged to form a star,  a gravitationally bound system.  Before the merger, these particles expanded with the FLRW spacetime until a time ${\tau_{\rm{star}}}$ when it decoupled from the  Hubble flow. One can write the action(equation \eqref{eq:massive_particle_actionl}) for the Hydrogen atom as
\begin{eqnarray}\label{eq:lightelement}
S_{\rm{H}} (\gamma_{\pm},\gamma'_{\pm}) &=& \int_{\tau_{\rm{ini}}}^{\tau_{\rm{star}}}
 {L} _{\rm{H}+}\left[ \gamma_{+}( \tau_{+}) , \gamma'_{+}( \tau_{+})\right] \d  \tau_{+} 
 {+}\int_{\tau_{\rm{star}}}^{\tau_{\rm{today}}}
  {L} _{\rm{H}-}\left[ \gamma_{-}( \tau_{-}) , \gamma'_{-}( \tau_{-})\right] \d  \tau_{-}\,,
\end{eqnarray}
where $ {L} _{\rm{H}+}$ is the lagrangain of particles of mass $m_{H}$ on $g_{ab}^{+}$ and $ {L} _{\rm{H}-}$ is the corresponding action $g_{ab}^{-}$.
As the universe expands, the one-parameter family of geodesics of light elements(a cloud of gas) reaches a critical time defined by  $\Theta =0$ and decouples from the Hubble flow.  This may be related to the  Jeans length, but we do not pursue the connection at the moment.  At $\Theta =0$, the local region defined by the matter horizon $r < r_{\rm{MH}}$, lose their global FLRW  identity.  The subsequent evolution of particles within this region takes place in a separate universe with coordinate time flowing backwards.

\begin{figure}[h]
 \includegraphics[width=100mm,height=90mm]{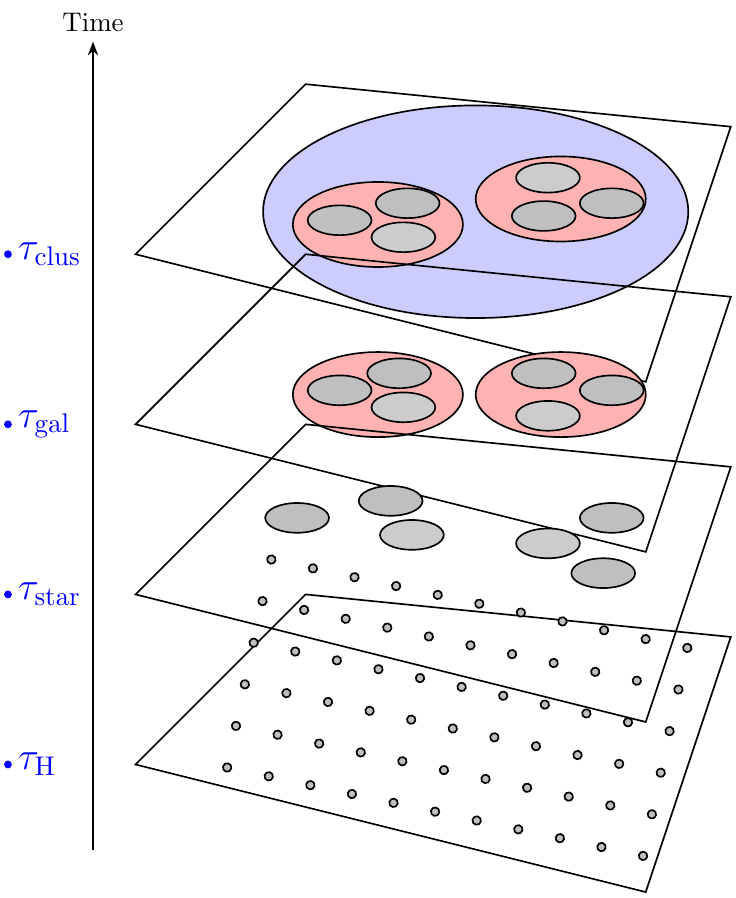}
 \caption{
 This is an illustration of how light elements(particles), through gravitational collapse instability,  form massive particles after some time has elapsed. The left vertical axis indicates the flow of proper time,  The left vertical line has crucial time scales:  $\tau_{ \rm{H}}$, $\tau_{ \rm{star}}$,  $\tau_{ \rm{gal}}$ and $\tau_{ \rm{clus}}$ which denote maximal hypersurfaces for the hierrachy of mass of corresponding to gravitational bound system, for example: starting from the Hydrogen atom, to star, galaxy and cluster.  
}
 \label{fig:obs_multi_scale}
\end{figure}

\item {\tt{Star formation scale}}

The  decoupling from the Hubble flow by a family of individual light elements leads to the formation of stars with mass:
$m_{\rm{star}} = \sum_{i = 1}^{N} m^{\rm{H}}_{i}  \,,$
where $m^{\rm{H}}_i$  is the mass of the light element.
The star is a gravitationally bound massive particle; hence, its propagation follows a time-like geodesic as well
\begin{eqnarray}
S_{\rm{Star}} (\gamma_{\pm},\gamma'_{\pm}) &=& \int_{\tau_{\rm{Star}}}^{\tau_{\rm{Gal}}}
 {L} _{\rm{star}+}\left[ \gamma_{+}( \tau_{+}) , \gamma'_{+}( \tau_{+})\right] \d  \tau_{+} 
{+}\int_{\tau_{\rm{Gal}}}^{\tau_{\rm{today}}}
  {L} _{\rm{star}-}\left[ \gamma_{-}( \tau_{-}) , \gamma'_{-}( \tau_{-})\right] \d  \tau_{-}\,,
\end{eqnarray}
Unlike the initial conditions for the geodesics of the light elements which may be determined by the physics of nucleosynthesis, the initial conditions for the geodesics of $m_{\rm{star}}$ are determined by the maximal foliation condition $\Theta = 0$ and the total momentum of the system of gravitationally bound particles $
{\bf{v}}_{\rm{star}} =
 \sum_{i=1}^{ N} m^{\rm{H}}_i \star {\bf{v}}^i_{\rm{H}}/m_{\rm{star}} \,,
$
where  ${\bf{v}}_{\rm{star}}$ is the velocity of the center of mass of the entire particles within the gravitationally bound star and ${\bf{v}}^i_{\rm{H}}$ is the velocity of the i-th particle.  
The proper time of the observer at rest in $m_{\rm{star}}$  at $\tau_{\rm{star}}$  synchronised to the forward-flowing FLRW spacetime. 
The effect of gravity is to pull massive particles together; stars are attracted to each other until $\Theta =0$, when the local sub-region decouples from the Hubble flow. forming a gravitationally bound galaxy. The individual stars making up the galaxy lose their global identity; then they evolve as a separate universe with coordinate time flowing backwards.
.

\item {\tt{Galaxy formation scale}}

Similarly, galaxies form from collections of stars, etc,  bound together by gravity: $m_{\rm{gal}} = \sum_{i = 1}^{N} m^{\rm{star}}_{i} $, 
where $m^{\rm{star}}_{i}$ is the mass of individual stars. 
The action for the geodesic is given by
\begin{eqnarray}
S_{\rm{gal}} (\gamma_{\pm},\gamma'_{\pm}) &=& \int_{\tau_{\rm{gal}}}^{\tau_{\rm{clus}}} 
 {L} _{\rm{gal}+}\left[ \gamma_{+}( \tau_{+}) , \gamma'_{+}( \tau_{+})\right] \d  \tau_{+}
{+}
\int_{\tau_{\rm{clus}}}^{\tau_{\rm{today}}}
 {L} _{\rm{gal}-}\left[ \gamma_{-}( \tau_{-}) , \gamma'_{-}( \tau_{-})\right] \d  \tau_{-} \,,
\end{eqnarray}
The first segment describes the propagation of galaxies in the expanding spacetime with forward time orientation. A collection of galaxies decouples from the Hubble flow when the rate of growth of the local overdensity is proportional to the Hubble rate, forming a gravitationally bound cluster at about $\tau_{\rm{clus}} $. The initial velocity of the galaxy, determined from the momentum conservation, is given by
$
{\bf{v}}_{\rm{gal}} = \sum_{i=1}^{ N} m^{\rm{star}}_i \star {\bf{v}}^i_{\rm{star}}/m_{\rm{gal}}\,,
$
where  ${\bf{v}}_{\rm{gal}}$ is the velocity of the centre of mass of the gravitationally bound star and ${\bf{v}}^i_{\rm{star}}$ is the velocity of the i-th star.  Further details on this are shown in Figure \ref{fig:obs_multi_scale}.

\item {\tt{Cluster formation scale:}}

The same mechanism applies in the case of a cluster. Here, a collection of galaxies mainly determines the mass of a cluster.
$ m_{\rm{clust}} = \sum_{i = 1}^{N} m^{\rm{gal}}_{i} $
where $m^{\rm{gal}}_{i}$ is the mass of individual galaxies.
The action for the cluster geodesic is given by
\begin{eqnarray}
S_{\rm{clus}} (\gamma_{-},\gamma'_{-}) &=& \int_{\tau_{\rm{clus}}}^{\tau_{\rm{sup. clus}}}
{L} _{\rm{clus}}\left[ \gamma_{+}( \tau_{+}) , \gamma'_{+}( \tau_{+})\right] \d  \tau_{+} 
{+}\int_{\tau_{\rm{sup. clus}}}^{\tau_{\rm{today}}} 
 {L} _{\rm{clus}-}\left[ \gamma_{-}( \tau_{-}) , \gamma'_{-}( \tau_{-})\right] \d  \tau_{-}
\end{eqnarray}
The initial velocity is given by the centre of mass velocity of the collection of galaxies
$
{\bf{v}}_{\rm{clus}} =\sum_{i=1}^{ N} m^{\rm{gal}}_i \star {\bf{v}}^i_{\rm{Gal}}/m_{\rm{clust}} \,,
$
where  ${\bf{v}}_{\rm{clus}}$ is the velocity of the centre of mass of the entire galaxy within the cluster.
Although most clusters live in superclusters, superclusters themselves are loosely bound by gravity at the moment. In the future, this will change as the expanding universe with forward-flowing coordinate time becomes less dense.

\end{itemize}

Finally, the multi-scale structure formation described in figure \ref{fig:obs_multi_scale} is consistent with the picture presented in figure \ref{fig:resolution} for the cosmological zoom-in N-body simulation approach for probing a small-scale region at high resolution.

\section{Conclusion}\label{sec:conc}

We have demonstrated that the breakdown of the predictive control of perturbative descriptions for large-scale structure clustering, which occurs at shell crossing, is not an abrupt failure but is preceded by a physically and geometrically significant event: the formation of a matter horizon. For inhomogeneities evolving on an expanding FLRW background spacetime, we have shown that the matter horizon always forms prior to the development of caustics. This finding provides a new and robust diagnostic for identifying the limits of perturbative methods.

The concept of a matter horizon was initially introduced by Ellis and Stoeger in 2010~\cite{Ellis:2010fr}  as a dynamical causal boundary delineating the region from which matter has been gravitationally drawn to form a specific structure. Our work formalises this concept within a general relativistic framework, establishing it as a key feature in the nonlinear evolution of cosmic structures.

To extend predictive control beyond the matter horizon and avoid the gravitational focusing singularity, we introduce a novel, multi-sheeted/scales spacetime description following Einstein and Rosen~\cite{Einstein:PhysRev.48.73}. This approach accounts for the change in the sign of the expansion scalar at the matter horizon, allowing us to describe the internal dynamics of the gravitationally bound region. We demonstrate that this internal region, which is characterised by faster timescales and matter trajectories that are decoupled from the global expanding FLRW spacetime, can be formally described as a separate, oriented spacetime with the coordinate time moving backwards.

This multi-scale framework provides a mathematically consistent and complete description of structure formation, bridging the gap between linear perturbation theory and the highly nonlinear regime. It offers a solution to the fundamental problem of how to evolve the spacetime description through shell crossing without resorting to unphysical singularities.

We further uncovered a deep correspondence between this relativistic multi-scale framework and the cosmological zoom-in N-body simulation approach. Specifically, the patch of spacetime enclosed by the matter horizon maps directly onto the region of interest in a zoom-in simulation, and both formalisms solve equivalent dynamical equations. In our construction, it is the internal domain of a gravitationally bound system, where dynamical timescales is short and it is treated as a separate universe with a reversed orientation of the coordinate time. This is conceptually identical to re-simulating a region of interest in cosmological zoom-in N-body simulations, with the matter horizon providing a natural and dynamically defined boundary for the region.

This work opens several avenues for future research. A primary goal is to quantitatively assess the backreaction of particle dynamics on the spacetime itself, an effect often neglected in standard cosmological models. Furthermore, the global symmetry we have identified can be used to define new quasi-local observables, which will provide a mathematically consistent framework for understanding complex phenomena such as galaxy bias and the interplay between structure formation and cosmic expansion. This work lays the groundwork for a more complete and predictive theory of large-scale structure formation.

  \section*{Acknowledgement}
  I benefited from discussions with Sravan Kumar,  Mathew Hall,  David Wands and Robson Christie.  I appreciate the support of the CIC Foundation; without them, this work would not have seen the light of day.

\section*{Data Availability}

The tensor algebraic computations in this paper were done with the open-source tensor algebra software, xPand (\url{https://github.com/Obinna/xPand})~\cite{Pitrou:2013hga}.  xPand is based on xPert(\url{http://www.xact.es/xPert/index.html})~\cite{Brizuela:2008ra}.
The numerical computations were done with Python \cite{van1995python} with extensive use of numpy~(\url{https://numpy.org/devdocs/user/building.html})~\cite{ harris2020array}.
We made use of COLOSSUS ({\url{https://bdiemer.bitbucket.io/colossus/}})~\cite{Diemer:2017bwl} to generate data for the dark matter halos. 
Plotting of results was done with matplotlib (\url{https://matplotlib.org/stable/users/project/citing.html}) \cite{Hunter:2007}.  
The code can be found here:: \url{https://github.com/Obinna/Essential_paper}


\providecommand{\href}[2]{#2}\begingroup\raggedright\endgroup

\end{document}